%% file: main.tex
\documentclass[12pt]{article}

\usepackage{url}
\usepackage{amsmath,amssymb,amsthm,amscd}
\usepackage{graphicx}
\usepackage{subcaption}

\usepackage{natbib}
\usepackage{algorithm}	
\usepackage{algorithmic}
\usepackage{bm}
\usepackage{bbm}
\usepackage{enumitem}

\usepackage{wrapfig}
\usepackage{lscape}
\usepackage{rotating}
\usepackage{epstopdf}

\usepackage{eurosym}
\usepackage{textcomp}
\usepackage{xcolor}
\usepackage{helvet}
\usepackage{tikz}

\usepackage{hyperref}
\usepackage{cleveref}
\hypersetup{
   colorlinks=true,
   citecolor = black,
   linkcolor=black,   
   urlcolor=blue,
   }
\usepackage{cprotect}
\usepackage{booktabs}
\usepackage{array}   
\newcolumntype{C}{>{$}c<{$}} 

\usepackage[toc,page]{appendix}

\graphicspath{ {Images/} }

\usepackage{xr} 

\makeatletter

\theoremstyle{plain}

\newcommand{\ts}{^{\sf T}} 

\newcommand{\dif}[2]{\frac{{\rm d} #1}{{\rm d} #2}}

\newcommand{\ilpdif}[2]{\partial #1/{\partial #2 }}
\newcommand{\pdif}[2]{\frac{\partial #1}{\partial #2}}
\newcommand{\pddif}[3]{\frac{\partial^2 #1}{\partial #2 \partial #3}}
\newcommand{\ptdif}[4]{\frac{\partial^3 #1}{\partial #2 \partial #3 \partial #4}}

\newcommand{\diag}[1]{\operatorname{diag}\left(#1\right)}
\newcommand{\tr}[1]{\operatorname{tr}\left(#1\right)}				

\newcommand{\var}{\operatorname{var}}

\newcommand*\rel@kern[1]{\kern#1\dimexpr\macc@kerna}
\newcommand*\widebar[1]{%
  \begingroup
  \def\mathaccent##1##2{%
    \rel@kern{0.8}%
    \overline{\rel@kern{-0.8}\macc@nucleus\rel@kern{0.2}}%
    \rel@kern{-0.2}%
  }%
  \macc@depth\@ne
  \let\math@bgroup\@empty \let\math@egroup\macc@set@skewchar
  \mathsurround\z@ \frozen@everymath{\mathgroup\macc@group\relax}%
  \macc@set@skewchar\relax
  \let\mathaccentV\macc@nested@a
  \macc@nested@a\relax111{#1}%
  \endgroup
}

\newcommand{\subR}[4][0.02, 0.95]{%
    \begin{tikzpicture}
        \node[anchor=south west, inner sep=0] (image) at (0,0) {
            \includegraphics[width=#2]{#3}
        };
        \begin{scope}[x={(image.south east)}, y={(image.north west)}]
            \node[anchor=north west, font=\sffamily\small] at (#1) {#4};
        \end{scope}
    \end{tikzpicture}%
}

\newcommand{\blind}{1}

\newcommand{\papertitle}{Integrating Complex Covariate Transformations in Generalized Additive Models}
\newcommand{\paperauthors}{Claudia Collarin, \\
    School of Mathematics, University of Edinburgh, \vspace{0.4cm} \\ 
    Matteo Fasiolo, \\
    \vspace{0.4cm}
    School of Mathematics, University of Bristol, \\ 
    Yannig Goude, \\
    \vspace{0.4cm}
    {\'E}lectricit{\'e} de France R\&D, \\
    Simon N. Wood \\
    School of Mathematics, University of Edinburgh}

\begin{document}

\def\spacingset#1{\renewcommand{\baselinestretch}%
{#1}\small\normalsize} \spacingset{1}


\if1\blind
{
  \title{\bf \papertitle}
 \author{\paperauthors}
    \date{}
  \maketitle
} \fi

\if0\blind
{
  \bigskip
  \bigskip
  \bigskip
  \begin{center}
    {\LARGE\bf \papertitle}
\end{center}
  \medskip
} \fi

\bigskip
\begin{abstract}
Transformations of covariates are widely used in applied statistics to improve interpretability and to satisfy assumptions required for valid inference. More broadly, feature engineering encompasses a wider set of practices aimed at enhancing predictive performance, and is typically performed as part of a data pre-processing step. In contrast, this paper integrates a substantial component of the feature engineering process directly into the modelling stage. This is achieved by introducing a novel general framework for embedding interpretable covariate transformations within multi-parameter Generalised Additive Models (GAMs). Our framework accommodates any sufficiently differentiable scalar-valued transformation of potentially high-dimensional and complex covariates. These transformations are treated as integral model components, with their parameters estimated jointly with regression coefficients via maximum a posteriori (MAP) methods, and joint  uncertainty quantified via approximate Bayesian techniques. Smoothing parameters are selected in an empirical Bayes framework using a Laplace approximation to the marginal likelihood, supported by efficient  computation based on  implicit differentiation methods. We demonstrate the flexibility and practical value of the proposed methodology through applications to forecasting electricity net-demand in Great Britain and to modelling house prices in London.  
\if1\blind{
Methods for building and fitting GAMs with nested transformations are provided by the gamFactory R package, available at \url{https://github.com/mfasiolo/gamFactory}, while the code for reproducing the results in this paper is available at \url{https://doi.org/10.5281/zenodo.19239350}.
}\fi
\end{abstract}


\noindent%



\section{Introduction} \label{sec:introduction}

Transformations of independent variables are a standard tool in applied statistics. They involve modifying covariates to enhance the interpretability of statistical models or to satisfy assumptions that are essential for valid inference. While covariate transformations focus on altering existing variables, feature engineering encompasses a broader set of practices \citep{verdonck2024special}, including the creation of new variables and the application of dimension reduction techniques, and typically places greater emphasis on predictive performance. In most cases, both variable transformations and feature engineering are carried out as part of the pre-processing stage, that is, prior to fitting the chosen model. 

In contrast, the present work aims to fully integrate part of the feature engineering process into the modelling phase. We focus particularly on interpretable transformations designed to handle complex covariates, such as time series and spatial data, and on their embedding into multi-parameter generalised additive models (GAMs), which include generalised additive models for location, scale and shape (GAMLSS; \citealp{rigby2005generalized}) as a special case. In particular, we extend GAMs to accommodate smooth effects that incorporate any scalar-valued, nested covariate transformation that is sufficiently differentiable with respect to (w.r.t.) its own parameters. These transformations are treated as integral components of the model, with their parameters estimated jointly with the regression coefficients using maximum a posteriori (MAP) methods. Joint uncertainty estimates are obtained via approximate Bayesian methods. Smoothing parameters are selected within an empirical Bayes framework by maximising a Laplace approximation to the marginal likelihood (LAML) using quasi-Newton optimisation. As this procedure requires evaluation of the LAML gradient, we provide efficient methods for computing it by extending the implicit differentiation techniques of \citet{wood_smoothing_2016} to exploit the specific structure of the models considered here. In addition, we propose a principled and theoretically well-founded solution to the scaling problem that arises when constructing spline-based smooth effects of parameter-dependent covariate transformations.

\begin{figure}[!t]\centering
    \begin{subfigure}{\textwidth}\centering
        \raisebox{2pt}{\subR[0.13, 0.855]{0.31\textwidth}{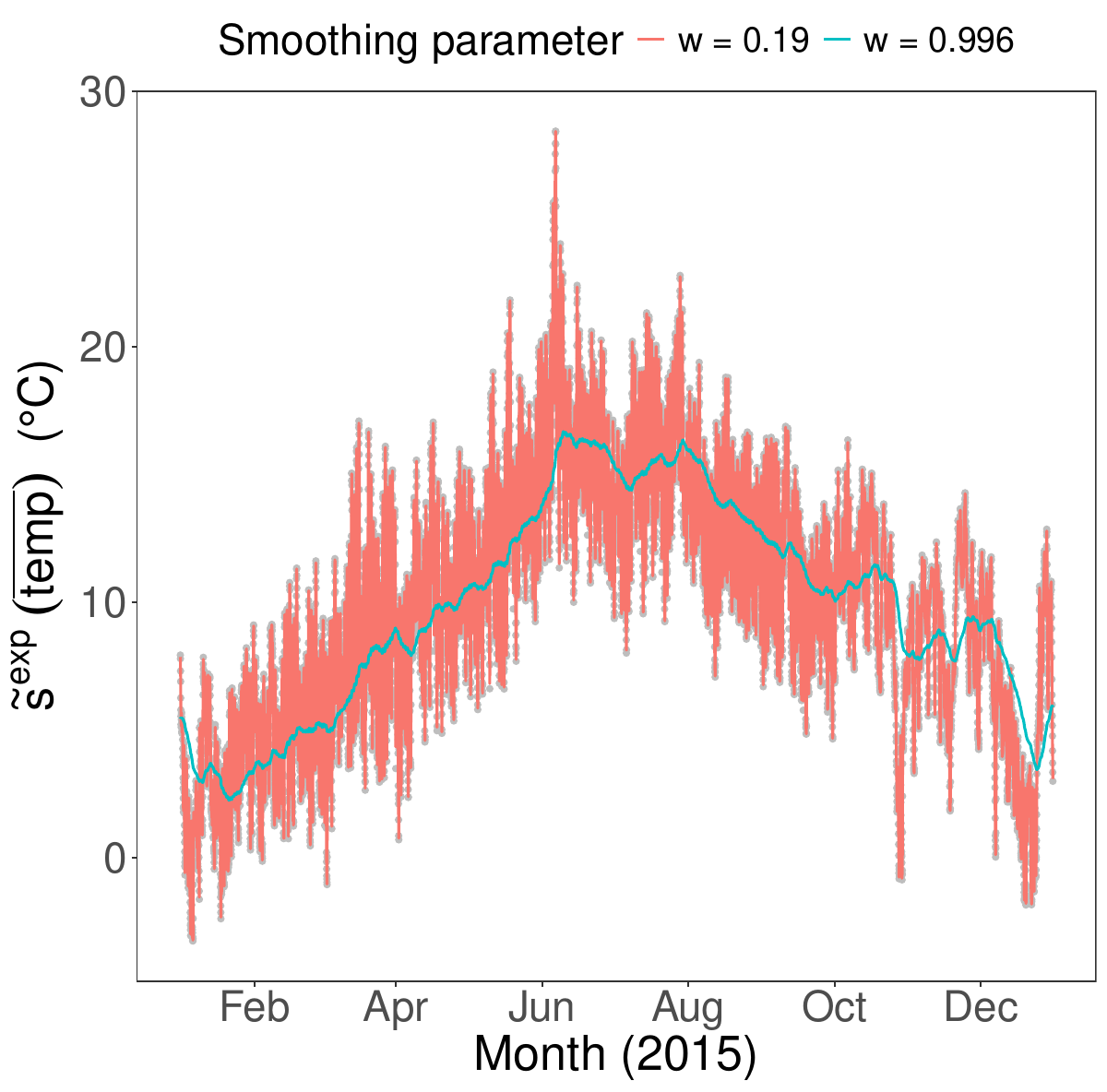}{a}}
        \subR[0.87, 0.87]{0.31\textwidth}{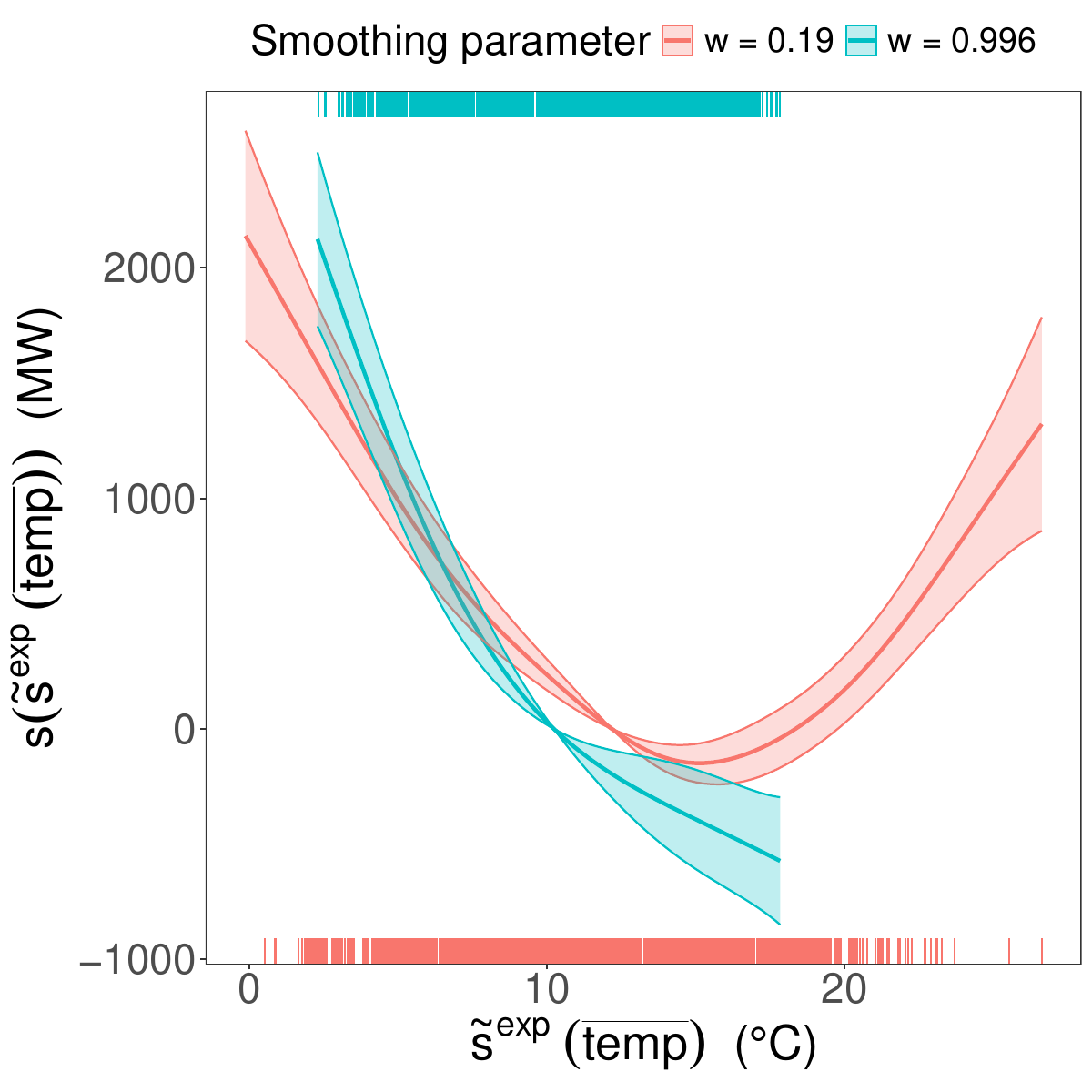}{b}
        \raisebox{-0.5pt}{\subR[0.13, 0.88]{0.31\textwidth}{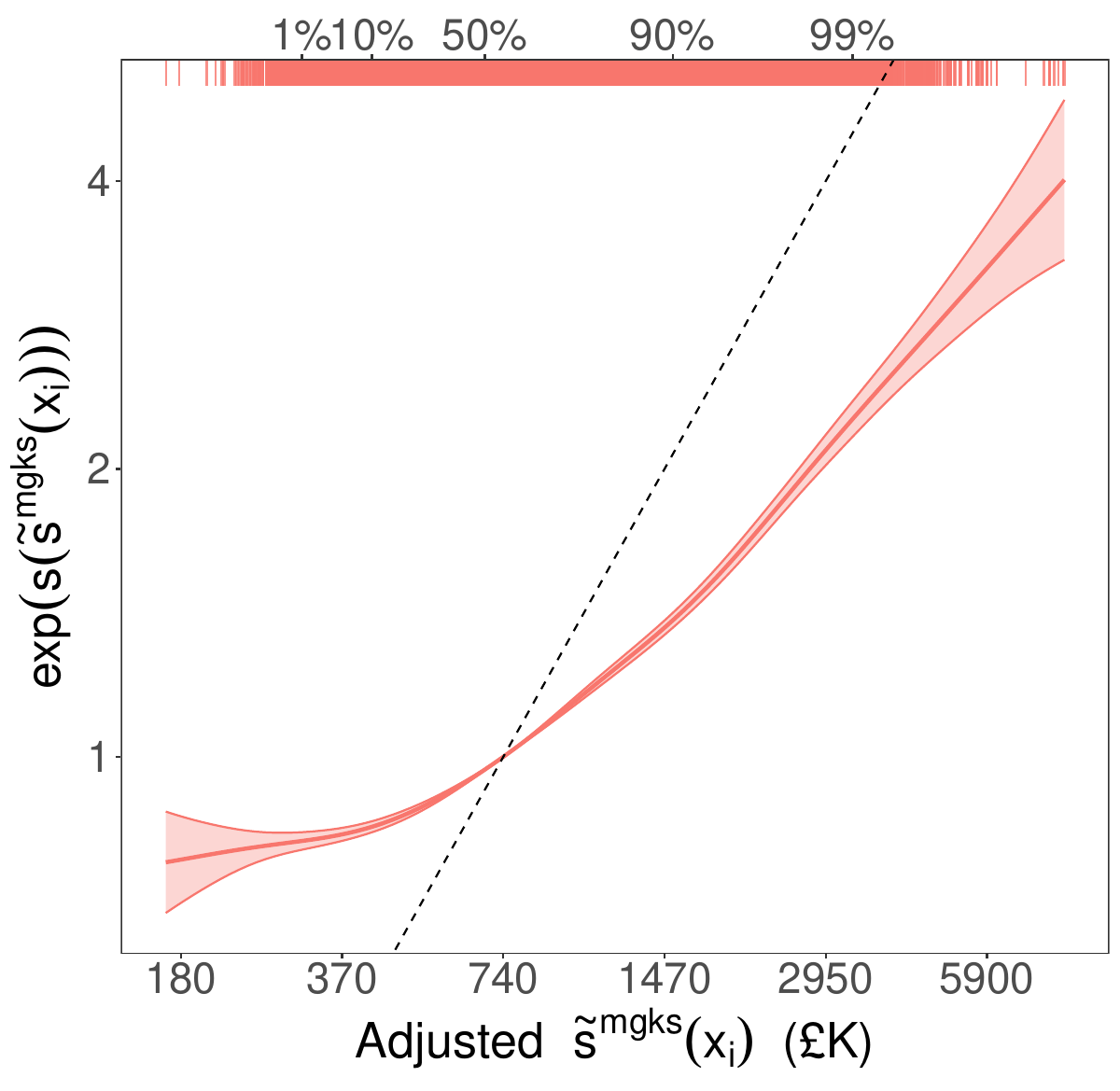}{c}}
    \end{subfigure}\\
    \begin{subfigure}{\textwidth}\centering
         \subR{0.455\textwidth}{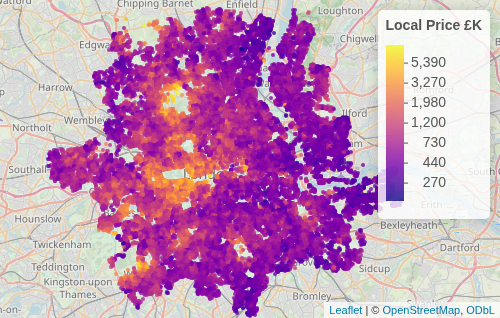}{d}
         \qquad
         \subR{0.455\textwidth}{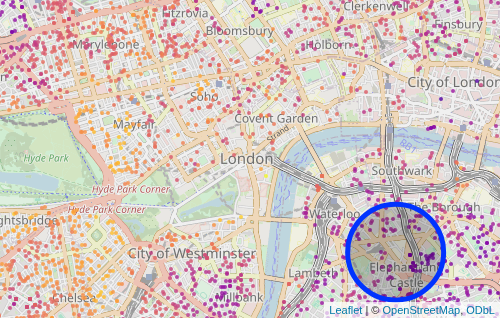}{e}
    \end{subfigure}
    \caption{Examples of smooth effects with nested covariate transformations. Exponential smooths of forecast temperature in GB (a) and the corresponding smooth effects on electricity net-demand (b) for two distinct exponential smoothing parameters, estimated by model \eqref{eq:netdemand}. Local house prices in London (d) and central London (e), obtained by kernel smoothing. The radius of the blue circle ($\approx 520$ meters) is twice the bandwidth of the kernel estimated by model \eqref{eq:mod_house}. The estimated multiplicative effect of kernel-smoothed prices on local expected house prices is shown in plot (c). The reference dashed line has unit slope and the five ticks at the top are price quantiles. }\label{fig:first_examples}
\end{figure}

The model class proposed here is widely applicable, but was initially motivated by problems in forecasting electricity net-demand, that is, consumption minus embedded generation. Consider the problem of forecasting the total hourly net-demand, $y_t$, in Great Britain (GB) using, among other covariates, a hourly forecast of external temperature, $\text{temp}_t$. Owing to buildings' thermal inertia, the consumption due to electrical heating and cooling at time $t$ is not entirely driven by $\text{temp}_t$, but depends on $\text{temp}_{t-1}$, $\text{temp}_{t-2}$, $\dots$, as well.        
One way to capture thermal inertia is to include in the forecasting model an exponentially smoothed temperature covariate, $\text{temp}_t^S$. An example is provided by Figure~\ref{fig:first_examples} (a-b), which shows two smoothed temperature trajectories (a) and their effects on the expected net-demand (b). Here, estimating the exponential smoothing parameter during model fitting, rather than via expert knowledge, allowed us to identify a temperature effect characterised by low inertia (red), related to both heating and cooling, and a smoother one (blue), related to heating only. The effects correspond to a model described in Section \ref{sec:application_netdemand}.

As a further example, consider a London house prices modelling application, which will be described in detail in Section~\ref{sec:application_house}. Spatial residual autocorrelation is often strong in house prices due to unobservable local characteristics, events such as gentrification and foreclosures, as well as price spillover, that is, the effect of recently observed
sale prices in the neighbourhood \citep{kallberg_crime_2025, guerrieri2013endogenous, fischer2018spillover}. 
Spatial autoregressive models capture such effects by first weighting neighbouring sale prices to build a local price index, which is then considered fixed when fitting the regression model \citep{lesage2009introduction}. The methods proposed here allow to simultaneously select the rate of decay of the spatial weights and estimate the non-linear effect of the local weighted price index on expected prices. An example is provided by Figure~\ref{fig:first_examples} (c-e), which shows the estimated local log-price index (d-e) and its effect on expected log-price (c).

To the best of our knowledge, the work proposed here is the first to fully integrate general covariate transformations within a GAM modelling framework. Nevertheless, covariate transformations have a long-standing history in statistics, and the automatic estimation of transformation parameters was first proposed by \citet{box_1964}. 
However, their work, and many subsequent developments such as \cite{thompson_transformations_2003} and \cite{fan_linear_2013}, have focussed on finding transformations that linearise the relation between the response and the covariates. In contrast, here we flexibly capture non-linearities via smooth effects, and use nested covariate transformations to integrate the processing of complex covariates into the model.

From this perspective, functional GAMs \citep[FGAMs,][]{mclean_functional_2014, greven2017general} provide an alternative to the methods proposed here in certain applications. Specifically, they provide GAM methods meant to incorporate functional covariates within the model, thus avoiding the use of a pre-processing step aimed at summarising them to scalar covariates. Similarly, penalised distributed lag effects \citep{zanobetti2000generalized, muggeo_modeling_2008, gasparrini_penalized_2017} represent an alternative to smooth effects of linear combinations of covariates when capturing the effect of several lagged values of the same explanatory variable. These effects are a special case of the class of transformations considered here and will be referred to as \textit{single index effects}. Further, latent Gaussian models \citep{rue2009approximate, lindgren2011explicit} can be used in place of the nested transformation provided here to capture spatio-temporal dependencies and correlations. 


The model class proposed here is also closely related to projection pursuit regression \citep[PPR,][]{friedman1981projection, collins2024bayesian} and (generalised) partially linear single index models \citep[GPLISM,][]{carroll1997generalized, antoniadis2004bayesian,yu_penalised_2017
}. In particular, both model classes include single index effects. 
Hence, the proposed framework extends such models by offering a wider range of transformations, as well as by allowing for multiple linear predictors and response distributions beyond the exponential family.

The rest of the article is structured as follows. Section~\ref{sec:method} introduces the proposed model structure, with three examples of covariate transformations, and gives details on how to address the scaling issue mentioned above. Section~\ref{sec:model_fitting} focusses on the fitting, computational and inferential framework, while Section~\ref{sec:applications} considers applications to net-demand forecasting in GB and London house prices modelling. 

\section{Integrating covariate transformations in GAMs}
\label{sec:method}

\subsection{The general model structure}
\label{sec:model_struct}
Let ${\mathbf y} = \{y_1, \dots, y_n\}$ be a vector of response variables and $\mathbf{x}_1, \dots \mathbf{x}_n$ the corresponding $d$-dimensional vectors of covariates. Assume that the $y_i$'s are conditionally independent given $\mathbf{x}_i$ and follow the distribution $\mathcal{D}(y_i; {\bm \theta}_i)$, with probability density function (p.d.f.) $p(y_i|\mathbf{x}_i)$.  In what follows, a \textit{covariate transformation} is a scalar-valued function $\tilde{s}:\mathbb{R}^p \to \mathbb{R}$ parameterised by a vector $\bm{a}$ and fourth-order differentiable w.r.t. the latter. A \emph{nested smooth effect}, $s$, is a fourth-order differentiable smooth function whose argument is a covariate transformation, that is, $s[\tilde{s}(\cdot)]$. The elements of the parameter vector $\bm{\theta}_i$ are modelled by
\begin{equation} \label{eq:basicGAM}
    g_j(\theta_{j i})
        = \eta_{ji} = \left(\mathbf{Z}^0_{ji}\right)\ts \bm{\gamma}_{j 0} 
        + \sum_{k=1}^{K_j} f_{j k}\left(\mathbf{x}^{S_{j k}}_{i}\right) 
        + \sum_{u=1}^{U_j} s_{j u}\left[\tilde{s}_{j u}\left(\mathbf{x}^{\tilde{S}_{j u}}_{i}\right)\right], 
        \qquad \text{for} \qquad j = 1, \dots, m.
\end{equation} 
where $g_j$ is a monotonic function, $\mathbf{Z}^0_{ji}$ is the $i$-th row of the design matrix $\mathbf{Z}^0_j$, and $\bm \gamma_{j0}$ is a vector of regression coefficients. We indicate standard smooth terms with $f_{jk}(\cdot)$ and nested smooth effects with $s_{j u}\left[\tilde{s}_{j u}\left(\cdot\right)\right]$. Coefficients associated with a given term inherit the same superscripts. For example, ${\bm a}^{ju}$ represents the vector that parameterises the covariate transformation $\tilde{s}_{ju}$. The covariates that each smooth depends on are denoted by $\mathbf{x}_i^{S_{ju}}$, where, $S_{ju} \subseteq \{1, \dots, d\}$. For example, if $S_{ju} = \{1, 3\}$, then $\mathbf{x}_i^{S_{ju}}$ is a two-dimensional vector consisting of the first and third elements of $\mathbf{x}_i$. Each function $f_{jk}$ is built via
\begin{equation}\label{eq:spline_rep}
    f_{jk}\left(\mathbf{x}_i^{S_{jk}}\right) = \sum_{l=1}^{L_{jk}} h^{jk}_l \left(\mathbf{x}_i^{S_{jk}}\right) \gamma_l^{jk}
\end{equation}
where the $h^{jk}_l$'s are basis functions of dimension $\vert S_{jk} \vert$ and the $\gamma_l^{jk}$'s are regression coefficients. 
Each $s_{ju}$ is built via a linear combination of 
basis functions, as in \eqref{eq:spline_rep}, with the additional requirement that the basis functions must be fourth-order differentiable. This is imposed by the efficient fitting framework described in Section~\ref{sec:model_fitting}. While the argument of $s_{ju}$ depends on ${\bm a}^{ju}$, the basis functions themselves (e.g., the position of the knots) should not depend on ${\bm a}^{ju}$. See Section \ref{sec:scaling_issues} for more details. Any basis function that meets the above requirements can be used to construct ${s}_{ju}$, and there are no restrictions for $f_{jk}$. 

The number of basis functions used to build each $f_{jk}$ and $s_{ju}$ is chosen to be large enough to avoid over-smoothing. The wiggliness of these effects is controlled by an improper multivariate Gaussian prior on the regression coefficients, which is centred at zero and shrinks the effects toward smoothness. The definition of smoothness is determined by the specific effect and the associated prior. This is implemented through positive semidefinite matrices, $\mathbf{S}_g$, with dimensions matching the total number of model coefficients. Each matrix $\mathbf{S}_g$ is sparse, containing non-zero entries only for the specific coefficients it penalises, while the remaining elements are padded with zeros. The prior precision matrix is ${\bf S}^{\bm \lambda} = \sum_{g=1}^G \lambda_g {\bf S}_g$, where $G$ is the total number of penalty matrices and 
$\bm \lambda = \{\lambda_1, ..., \lambda_G\}$ is a vector of positive smoothing parameters. The prior can also include the parameters of the $\tilde{s}_{ju}$'s. However, its interpretation depends on the type of transformation and prior considered.

\subsection{Instances of covariate transformations} \label{sec:transformations}
Here we detail three instances of nested transformation, namely adaptive exponential smoothing, multivariate kernel smoothing and linear combinations.


\subsubsection{Adaptive exponential smoothing} \label{sec:exp_smooth}

Let $x_i$ be the $i$-th observed value of a scalar covariate. An adaptive exponential smoothing transformation is 
\begin{equation}\label{eq:exp_smooth}
    \tilde{s}(x_i) = \tilde{s}_i = 
       \omega_{i}\tilde{s}_{i-1}+(1-\omega_{i})x_{i}, \qquad \text{for} \qquad  i \geq 1, 
\end{equation}
where $\tilde{s}_0 = x_0$ and $\omega_i \in (0, 1)$. The smoothing factor can be modelled via $\omega_i = \phi({\tilde{\bf  x}}_i\ts {\bm a})$, where $\phi$ is the logistic function, $\tilde{\bf x}_i$ is a fixed vector, and ${\bm a}$ is a vector of parameters. In Section~\ref{sec:application_netdemand} we provide an example where $\omega_i$ depends on $\tilde{\mathbf{x}}_i\ts = [1, \Delta h_i - 1]$, with $\Delta h_i$ being the time interval in hours between observations $i$ and $i-1$. In principle, $\omega_i$ could be either constant or modelled via a full additive model, involving the sum of several parametric and penalised, non-parametric effects. The logistic function is used to ensure that $\omega_i \in (0, 1)$, but in principle any unconstrained, fourth-order differentiable parametrisation could be used.
Figure~\ref{fig:first_examples}a shows an example of an exponential smoothing transformation. 

\subsubsection{Multivariate kernel smoothing} \label{sec:kern_smooth}

Let ${z}_{i}$ be a scalar covariate corresponding to the vector ${\bf x}_{i}$. For example, ${z}_{i}$ might be the temperature measured at the location ${\bf x}_{i}$.
A kernel smooth of $z$, evaluated at ${\bf x}_i$, is
\begin{equation}\label{eq:mgks}
    \tilde{s}({\bf x}_i) = \frac{\sum_{j\in \mathcal{N}_i}K_{{\bm a}}({\bf x}_{i}, {{\bf x}}_{j})z_{j}}{\sum_{q\in \mathcal{N}_i}K_{{\bm a}}({\bf x}_{i},{{\bf x}}_{q})}, 
\end{equation}
where $K_{\bm a}$ is the kernel of a multivariate p.d.f., parametrised by vector $\bm a$ and $\mathcal{N}_i$ is the set of indices of ${\bf x}_i$'s neighbours. Note that $i$ might not appear in $\mathcal{N}_i$, as is the case for the kernel smoothing estimates of local house prices shown in Figure \ref{fig:first_examples}d-e. See Section \ref{sec:application_house} for a detailed description of the corresponding model. 

More generally, this transformation may be useful when dealing with spatially misaligned covariates. For example, one might wish to model the power production of a set of wind farms using wind speed measurements collected at several meteorological stations. Because the locations of the farms and the stations typically do not coincide, kernel smoothing provides a natural way to impute wind speeds at each farm. In a conventional two-step approach, the kernel bandwidth would first be selected using the weather station data alone, after which wind speeds would be interpolated at the farm locations to create an aligned data set for power modelling. In contrast, the methods proposed here allow the kernel bandwidth for wind speed to be selected directly by optimising the fit to the wind farms' power output.


In this work we consider the multivariate Gaussian kernel with bandwidth matrix $\bm \Sigma$ where $a_k = \log(1/\Sigma_{kk})$ and $\Sigma_{jk} = 0$ for $j \neq k$. We do not consider adaptive smoothing, but in principle this could be done by modelling the diagonal elements of $\bm \Sigma$ via an additive model, as we did for exponential smoothing. Further, one might consider modelling the full bandwidth matrix by, for example, using $\bm a$ to control its Cholesky factor. The fitting methods in Section \ref{sec:model_fitting} would support such a model, but we leave this to future work. 

\subsubsection{Linear combinations} \label{sec:linear_proj}

Let ${\bf x}_i$ be a vector of covariates; the linear combination
\begin{equation} \label{eq:si_model}
\tilde{s}({\bf x}_i) = {\bf x}_i\ts \bm a,
\end{equation}
allows for the inclusion of single index effects, i.e. smooth functions of linear combinations, into the model. Such transformations can be used to perform dimension reduction. Specifically, under the fitting framework proposed here, it is possible to specify interpretable multivariate Gaussian priors on the coefficient vector ${\bm a}$, which can be advantageous when ${\bm a}$ is high-dimensional. For example, in Section \ref{sec:application_netdemand} the elements of ${\bm a}$ are used to form a distributed lag effect, hence we use a prior penalising $\sum_k(a_k - a_{k-1})^2$, meant to encourage smoothness between the coefficients of consecutive lags. While we do not provide examples here, note that the elements of ${\bf x}_i$ could be the evaluated spline basis functions of some covariate $z_i$, i.e. ${\bf x}_i = {\bf x}(z_i)$, or more generally the $i$-th row of the model matrix of an additive model. That is, linear combinations can be used to build smooth effects whose argument is itself an additive model.  

From a computational perspective, linear combinations are a special case of transformations, because the linearity between $\tilde{s}$ and $\bm a$ makes the derivative system simple and more efficient to compute with, as explained in Section~\ref{sec:model_fitting}. However, there are simple non-linear variants of (\ref{eq:si_model}) that are compatible with the fitting framework provided here and could be considered in future work. For instance, $\tilde{s}$ could be a shape-constrained smooth effect, constructed using one of the non-linear parametrisations proposed by \cite{pya2015shape}. Alternatively, if $x_{ij}$ is a covariate measured at time $t_{i-j}$, with $j = 1, 2, \dots$, one might want to impose $|a_1| > |a_2| > \dots $, so that the weights of past covariates must decrease with the time lag. Such a constrained distributed lag effect could be implemented by adopting the parametrisation $a_1 = \tilde{a}_1$ and $a_j = a_{j-1}\phi(\tilde{a}_j)$, where each $\tilde{a}_j$ is unconstrained and $\phi:{\mathbb{R}}\rightarrow (-1, 1)$ is a monotonic and differentiable function.

\subsection{Nested smooth effect specification and identifiability} \label{sec:scaling_issues}

Integrating smooth effects with nested transformations within the empirical Bayes fitting framework adopted here presents additional challenges, relative to standard smooth effects. While in Section \ref{sec:model_fitting} we address the non-linearity of a nested effect, $s[\tilde{s}(\bf x)]$, w.r.t. the transformation's parameters, $\bm a$, here we focus on scaling and knots-placement issues. 

\subsubsection{The scaling problem}

Assume, for simplicity, that the outer smooth $s(\cdot)$ is built via a knot-based spline basis and note that the argument of $s(\cdot)$, i.e. $\tilde{s}(\bf x)$, depends on $\bm a$. Hence, as $\bm a$ is varied during model fitting, the spline basis functions are evaluated at different locations. For bases with finite support, such as B-splines, this creates a problem as there is no guarantee that all the values taken by $\tilde{s}(\bf x)$ fall within the support of $s(\cdot)$, as $\bm a$ varies. Also, the range of values taken by $\tilde{s}(\bf x)$, might be much smaller than the knot range of $s(\cdot)$, so that only a few basis functions contribute to the fit, leading to computational inefficiency and under-smoothing. This scaling problem is not limited to knot-based bases and can not be solved by adopting knot-free spline bases, such as thin plate splines. In fact, regardless of the type of basis being used, the scale and range of the covariate of interest must be taken into account when constructing the smooth effect basis. However, here the scale and range are $\bm a$-dependent, not fixed as for non-nested smooth effects. A na{\"i}ve solution to this scaling problem would be to rebuild the basis as $\bm a$ changes, for example by using knots that are equally-spaced across the range of values taken by $\tilde{s}(\bf x)$. However, doing so would likely compromise the differentiability of the likelihood function, thereby preventing the use of the efficient fitting framework proposed here.

\begin{figure}\centering
    \includegraphics[width = 0.8\textwidth]{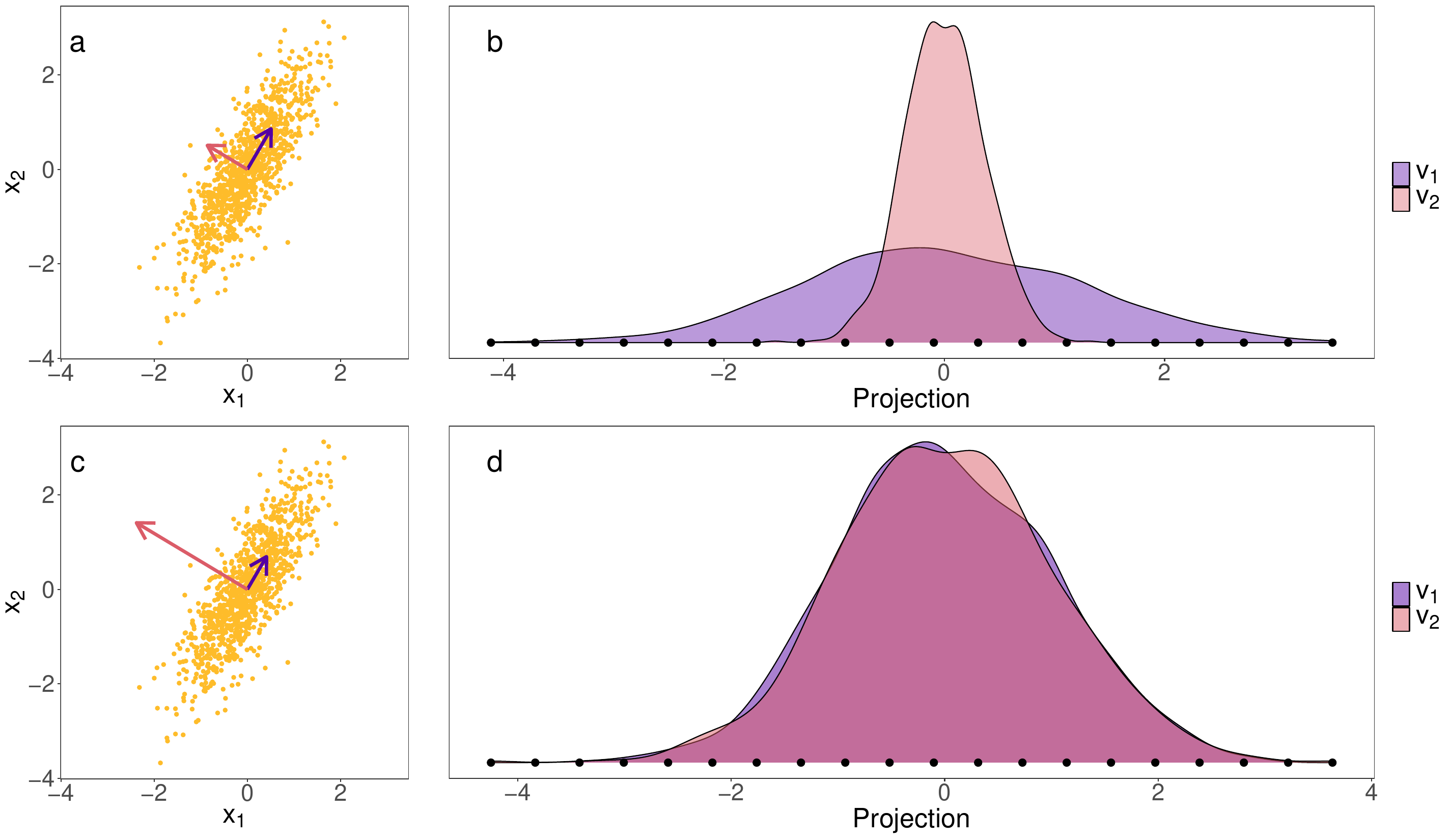}
    \caption{Scaling issues in linear combinations. Top row: Scatterplot plot of data to be combined via two unit-vectors (a) and densities of the transformed data (b). Bottom row: Same as on the top row, but here the norm of the vectors is adjusted so that the scale of the transformed data is invariant w.r.t. the direction of the combination vector.}\label{fig:scaling_issue}
\end{figure}

To get an intuition on the proposed solution, consider Figure~\ref{fig:scaling_issue}a-b. The scatterplot represents sample vectors ${\bf x}_i$, to be linearly combined via $\tilde{s}(\mathbf{x}_i)=\mathbf{x}_i\ts{\bm a}$, with $||\bm a|| = 1$. The densities on the right represent the distribution of the projected data and show that it depends on the direction of $\bm a$. Hence, a spline basis for the outer smooth $s(\cdot)$ with support on, say, $[-2, 2]$ would be too (wide) narrow when $\bm a$ is parallel to the (minor) major axis of the data ellipse. However, Figure~\ref{fig:scaling_issue}c-d shows that the scale of $\tilde{s}(\mathbf{x})$ can be made invariant by rescaling the norm of $\bm a$ based on its direction. 

This idea can be extended to a general transformation, $\tilde{s}(\mathbf{x})$, by imposing the constraints $\hat{\text{mean}}\left[\tilde{s}({\bf x})\right] = 0$ and $\hat{\var}\left[\tilde{s}({\bf x})\right] = 1$ on its sample mean and variance. While practical details on how to implement such constraints are provided in Supplementary Material \ref{app:scaling_penalty} (henceforth SM \ref{app:scaling_penalty}), below we explain how standardising the transformation helps to choose the range of knot-based outer smooth effects.

\subsubsection{Choosing the extreme knots}\label{sec:extreme_knots}
Indicate $\tilde{s}({\bf x}_i)$, where ${\bf x}_i$ is the $i$-th observed covariate value, simply with $\tilde{s}_i$, for $i = 1, \dots, n$. Similarly, use $\tilde{S}_{n+1} = \tilde{s}({\bf X}_{n+1})$ to indicate the transformation of a new, unobserved random covariate (e.g., from a test set). Assume that the sample mean and variance of $\tilde{s}_1,\ldots,\tilde{s}_n$ are fixed, respectively, to 0 and to some arbitrary positive constant $c$. 

Focussing on symmetric knot ranges of the type $\mathcal{R}(\xi) = [- \xi\sqrt{c}, \xi\sqrt{c}]$, we chose the value of $\xi > 0$ based on a deterministic upper bound, $\pi$, on the proportion of $s_i$'s that falls outside of $\mathcal{R}(\xi)$. In particular, assuming for simplicity that $\pi n/2$ is an integer, then the choice $\xi = \sqrt{(2-\pi)/\pi}$ guarantees that at most $\pi n$ of the observations will fall outside of $\mathcal{R}(\xi)$.
Further, the probability that a new transformed covariate, $\tilde{S}_{n+1}$, falls outside of $\mathcal{R}(\xi)$ can be upper bounded by 
\begin{equation*}
    \mathbb{P}\left[\tilde{S}_{n+1} \notin \mathcal{R}(\xi)\right]
	< \frac{1}{n+2}\left\lfloor\frac{n+2}{n+1}\left(\frac{n}{\xi^2}+1\right)\right\rfloor= \frac{\pi}{2-\pi} + O(n^{-1}), \;\;\; \text{with} \;\;\; \xi = \sqrt{\frac{2-\pi}{\pi}}.
\end{equation*}

The probabilistic bound involving $\tilde{S}_{n+1}$ is a direct application of the finite sample version of Chebyshev's inequality \citep{kaban_non-parametric_2012}. The deterministic bound is obtained by applying the extension of Samuelson's inequality obtained by \cite{wolkowicz_extensions_1979} which, given the constraints on the sample mean and variance, guarantees that the $k$-th order statistic of $(\tilde{s}_1, \ldots, \tilde{s}_n)$ falls in $[- \sqrt{c(n-k)/{k}}, \sqrt{c(k-1)/(n-k+1)}]$, for any $k=1,\ldots,n$. Setting $k = {\pi}n/2$, implies that at most $\pi n$ observations will fall outside the interval $ [-\sqrt{c(2-\pi)/\pi}, \sqrt{c(2-\pi)/\pi}]$. If $\pi n/2$ is non-integer, then setting $k = \lfloor\pi n/2\rfloor$ leads to at most $\lfloor \pi n\rfloor$ observations outside $\mathcal{R}(\xi)$.

Note that bounds provided above do not make any assumption on the distribution of the ${\bf x}_i$'s, hence they are fairly pessimistic. For example, in the applications discussed in Sections~\ref{sec:application_netdemand} and \ref{sec:application_house} we place the extreme knots using $\xi=6$ and $c = 1$, which leads to $\pi \approx 0.05$ and $ \pi/(2-\pi) \approx 0.025$. However, no transformed covariate in the in-sample and out-of-sample data falls outside $\mathcal{R}(\xi)$, under both models.  

\subsubsection{Constraints on the outer effect}
All the considerations discussed so far relate to the inner transformation $\tilde{s}$. Regarding the outer smooth effect, to ensure the identifiability of it, we imposed a set of point and derivative constraints. Specifically, the smooth was constrained to have no intercept, that is, $s(0) = 0$. This restriction avoids directly orthogonalising the smooth to the intercept term, i.e. $\sum_{j}s[\tilde{s}(x_j)]=0$, which would require the knowledge of $\tilde{s}({\bf x})$ before starting the fitting procedure. In addition, outside the support $\mathcal{R}(\xi)$, we imposed a linear extrapolation of the smooth effect. The definition of $\mathcal{R}(\xi)$ implies that during the optimisation procedure, a subset of $\tilde{s}_i$’s may still fall beyond the support. Therefore, to satisfy the continuity assumptions, we imposed higher-order derivative constraints at the boundary knots, requiring $s^{(2)}(\kappa) = s^{(3)}(\kappa) = s^{(4)}(\kappa) = 0$ for $\kappa \in \{-\xi\sqrt{c}, \xi\sqrt{c}\}$. 


\section{Model fitting and inference} \label{sec:model_fitting}

\subsection{Fitting framework overview} \label{sec:fitting_over}

Denote the vector containing all the regression coefficients and transformation parameters in the model with $\bm \zeta$ and the log-likelihood corresponding to the $i$-th observation with $\ell_i(\bm \zeta)$.
Under the Gaussian prior described in Section \ref{sec:model_struct} and the constraints on the sample variance of the nested transformations described in Section \ref{sec:scaling_issues}, the Bayesian posterior log-density can be expressed as
\begin{equation} \label{eq:logPosterior}
    \mathcal{L}(\bm \zeta)
    =\log p(\bm \zeta|{\bm y}, \bm \lambda) 
    = \sum_{i=1}^n \ell_i(\bm \zeta\,|\, y_i) - \frac{1}{2} \sum_{g=1}^G \lambda_g \bm \zeta \ts {\bf S}_g \bm \zeta, 
\end{equation}
up an additive constant. 
Under a multivariate Gaussian prior, the prior log-density is equivalent to a generalized ridge penalty. Consequently, high values of a smoothing parameter $\lambda_g$ lead to a posterior distribution that is more concentrated on the null space of the penalty. The definition of this null space depends on the choice of the prior precision matrix ${\bf S}_g$. Once ${\bf S}_g$ is chosen, it defines the concept of smoothness for the corresponding effect and the null space of ``completely smooth'' functions. Note that, in general, there is no one-to-one correspondence between the effects and the smoothing parameters. For instance, the wiggliness of an effect can be controlled using multiple smoothing parameters, while a single penalty can affect several effects.


For fixed smoothing parameters, $\bm \lambda$, we obtain MAP estimates of the regression coefficients by maximising the log-posterior (\ref{eq:logPosterior}) via Newton's algorithm. However, the main challenge is selecting the smoothing parameters themselves. We do this by maximising an approximation to the log-marginal likelihood, $\mathcal{V}(\bm \lambda) = \log p(\bm \lambda) =  \log \int p(\bm y | \bm \zeta)p(\bm \zeta|\bm \lambda) d \bm \zeta$. In particular, we consider a Laplace approximate marginal likelihood (LAML) criterion
\begin{equation} \label{eq:LAML}
\tilde{\mathcal{V}}(\bm \lambda) = \mathcal{L}(\hat{\bm \zeta})+\frac{1}{2}\log|{\bf S}^{\bm \lambda}|_{+}-\frac{1}{2}\log|\bm {\mathcal{H}}|+\frac{M_{p}}{2}\log(2\pi),
\end{equation}
where  $M_{p}$ is the dimension of the null space of ${\bf S}^{\lambda}$, $|{\bf S}^{\lambda}|_{+}$ is the product of its positive eigenvalues, $\bm{\mathcal{H}}$ is the negative Hessian of \eqref{eq:logPosterior}, evaluated at its maximiser, $\hat{\bm{\zeta}}$. To ensure the positivity of $\bm \lambda$, we maximise (\ref{eq:LAML}) w.r.t. $\bm \rho$, where $\rho_g = \log(\lambda_g)$. We use a BFGS optimiser, which requires the gradient of the objective
\begin{equation} \label{eq:LAML_grad}
(\nabla_{\bm \rho} \tilde{\mathcal{V}})_g = \frac{\partial \tilde{\mathcal{V}}}{\partial \rho_g} = - \frac{\lambda_g}{2} \hat{\bm \zeta}\ts {\bf S}_g \hat{\bm \zeta} + \frac{1}{2}\frac{\partial \log|{\bf S}^{\bm \lambda}|_{+}}{\partial \rho_g} - \frac{1}{2}\frac{\partial \log|\bm {\mathcal{H}}|}{\partial \rho_g}.
\end{equation}
Although computing the first two terms is straightforward, the third term is more involved. In particular, following \cite{wood_smoothing_2016}, we have
\begin{equation} \label{eq:DH_Drho}
\frac{\partial \log|\bm {\mathcal{H}}|}{\partial \rho_g} = \text{tr}\left(\bm {\mathcal{H}}^{-1} \frac{\partial \bm {\mathcal{H}}}{\partial \rho_g} \right).
\end{equation}
The direct calculation of \eqref{eq:DH_Drho} results in a computational cost of $O(np^3)$, where $p = \text{dim}({\bm \zeta})$, for each $\rho_g$. However, \cite{wood_smoothing_2016} shows how to achieve a more efficient $O(np^2)$ for standard GAMLSS models. Attaining similar computational efficiency for models with nested transformations is the key challenge in making the framework proposed here practically feasible for routine use. Section \ref{sec:deriv_framework} addresses this challenge.

\subsection{Efficient and modular derivative computation} \label{sec:deriv_framework}

Recall that maximising the log-posterior \eqref{eq:logPosterior} via Newton’s method requires the gradient and Hessian of the log-likelihood w.r.t. the regression coefficients. Expressions for these derivatives are provided in Section \ref{sec:llk_grad_hess}, while Section \ref{sec:D_hess_D_rho} focuses on the derivatives of the Hessian with respect to $\bm \rho$, which are required for gradient-based LAML maximisation. Deriving general expressions for these quantities while maintaining computational efficiency is not trivial, but the task is  facilitated by the notation introduced in Section \ref{sec:deriv_notation}.

\subsubsection{Setting up the notation}
\label{sec:deriv_notation}

The linear predictors in \eqref{eq:basicGAM} are modelled via a parametric component in addition to standard and nested smooth effects. We divide the nested effects into those based on linear combinations, which are a special case from a computational perspective, and the rest. Denote with ${\bm \alpha}$ the transformation parameters and with $\bm \beta$ the spline coefficients of an effect based on a linear combination. Define ${\bm a}$ and ${\bm b}$ similarly for a generic nested effect and let $\bm \gamma$ be the vector of coefficients belonging to a parametric and a standard smooth effects. 

Let ${\bf M}_{\bm b}$ and ${\bf M}_{\bm \beta}$ be the model matrices corresponding to the outer spline basis of the two classes of nested effects, and define ${\bf M}_{\bm \gamma}$ similarly for parametric or standard smooth effects. 
Denote with ${\bf M}_{\bm a}$ and ${\bf M}_{\bm \alpha}$ the matrices such that ${\bf M}_{\bm a} = \nabla_{\bm{a}}\ts{\tilde{\bf s}}$ and ${\bf M}_{\bm \alpha} = \nabla_{\bm \alpha}\ts{\tilde{\bf s}}$, where $\nabla_{\bm{a}}\ts\tilde{\bf s} = \ilpdif{\tilde{\bf s}}{\bm{a}\ts}$ and $\tilde{\bf s}$ is the vector containing the $n$ observed values of a transformation $\tilde{s}$. 
Note that, for linear combinations, ${\bf M}_{\bm \alpha} = {\bf X}^{\tilde{S}}$, where ${\bf X}^{\tilde{S}}$ is a matrix with $i$-th row ${\bf x}_i^{\tilde{S}}$. Indicate with ${\bf M}_{\bm b}^t$ and ${\bf M}_{\bm \beta}^t$ the matrices such that $({M}_{\bm b}^t)_{ik} = \partial^t ({M}_{\bm b})_{ik}/\partial \tilde{s}_i^t$ and $({M}_{\bm \beta}^t)_{ik} = \partial^t ({M}_{\bm \beta})_{ik}/\partial \tilde{s}_i^t$, with $t=1$ or 2. These are the derivatives of the outer spline bases of the nested effects w.r.t. the observed values of the transformation.

Let $\ell=\sum_{i}\ell_{i}$ be the log-likelihood and define the vector ${\boldsymbol \ell}$ such that its $i$-th element is $\ell_{i}$. Indicate with ${\boldsymbol \ell}^{\xi_1}$, ${\boldsymbol \ell}^{\xi_1\xi_2}$ and ${\boldsymbol \ell}^{\xi_1\xi_2\xi_3}$ the vectors such that $\ell_i^{\xi_1}=\partial{\ell}_i/\partial \xi_{i1}$, $\ell_i^{\xi_{i1}\xi_{i2}}=\partial^2{\ell}_i/\partial \xi_{i1}\partial\xi_{i2}$ and $\ell_i^{\xi_1\xi_2\xi_3}=\partial^3{\ell}_i/ \partial \xi_{i1}\partial\xi_{i2}\partial\xi_{i3}$, with $\xi_{ik}\in \left\{\theta_{ij}, \eta_{ij}, \tilde{s}_i\right\},\ k=1,2,3$. Note that the derivatives of $\ell_i$ w.r.t. $\theta_{ij}$ are model-specific, that is they depend on the response distribution, $\mathcal{D}(y_i; {\bm \theta}_i)$. Given these derivatives, and those of the inverse link function $g_j^{-1}$ w.r.t. $\theta_{ij}$, general expressions for the derivatives w.r.t. the linear predictor $\eta_{ij}$ and the transformation $\tilde{s}_i$ are readily obtained via the chain rule, as detailed in SM~\ref{app:D_ll_D_eta_s}.

Denote with $\bm \psi \in \{\bm \gamma,{\bm a}, {\bm b},  \bm \alpha, \bm \beta\}$ a generic vector of parameters and define \begin{equation*}
	\nu({\bm \psi}) = 
	\begin{cases}
		\eta_{\bm\psi} & {\bm \psi} \in \left\{\bm{\gamma}, \bm{\beta}, \bm{b}\right\}\\
		\tilde{s}_{\bm{\psi}} & \bm{\psi}\in \left\{\bm{\alpha}, \bm{a}\right\},
	\end{cases}
\end{equation*}
where the subscript indicates the linear predictor or the transformation that depends on $\bm \psi$. Hence, ${\boldsymbol \ell}^{\nu(\bm{\psi})}$ denotes ${\boldsymbol \ell}^{\eta_{\bm\psi}}$, if $\bm{\psi} \in \left\{\bm{\gamma}, \bm{\beta}, \bm{b}\right\}$,  or ${\boldsymbol \ell}^{\tilde{s}_{\bm{\psi}}}$, if $\bm{\psi} \in \left\{\bm{\alpha}, \bm{a}\right\}$. Higher-order derivatives follow the same convention. For example, ${\boldsymbol \ell}^{\nu(\bm{\psi}_1, \bm{\psi}_2)}$ indicates ${\boldsymbol \ell}^{\bm{\eta}_{\psi_1}\tilde{\bm{s}}_{\psi_2}}$, when $\bm{\psi}_1 \in \left\{\bm{\gamma}, \bm{\beta}, \bm{b}\right\}$ and $\bm{\psi}_2 \in \left\{\bm{\alpha}, \bm{a}\right\}$, and ${\boldsymbol \ell}^{\nu(\bm{\psi}_1, \bm{\psi}_2, \bm{\psi}_2)}$ denotes ${\boldsymbol \ell}^{\bm{\eta}_{\psi_1}\tilde{\bm{s}}_{\psi_2}\tilde{\bm{s}}_{\psi_3}}$, when $\bm{\psi}_1  \in \left\{\bm{\gamma}, \bm{\beta}, \bm{b}\right\} $ and $\bm{\psi}_2, \bm{\psi}_3 \in \left\{\bm{\alpha}, \bm{a}\right\}$. 

\subsubsection{Gradient and Hessian blocks of the log-likelihood} \label{sec:llk_grad_hess}

Under the notation just described, all the sub-vectors forming the gradient for the log-likelihood follow the general pattern $\nabla_{\bm{\psi}}\ell = {\bf M}_{\bm{\psi}}\ts{\boldsymbol \ell}^{\nu(\bm{\psi})}.$ In the Hessian matrix, most of the block types follow the pattern
\begin{equation} \label{eq:gen_Hess_pattern}
    \nabla_{\bm{\psi}_1}\ts\nabla_{\bm{\psi}_2}{\ell}  = 
	\pddif{\ell}{\bm{\psi}_1\ts}{\bm{\psi}_{2}} = {\bf M}_{\bm{\psi}_{2}}\ts{\bf H}^{\bm{\psi}_{1}\bm{\psi}_{2}}{\bf M}_{\bm{\psi}_{1}},
\end{equation}
where ${\bf H}^{\bm{\psi}_{1}\bm{\psi}_{2}}$ is a diagonal matrix
with non-zero elements 
$({\bf H}^{\bm{\psi}_{1}\bm{\psi}_{2}})_{ii}=\ell_{i}^{\nu(\bm{\psi}_{1},\bm{\psi}_{2})}$. The only block types that do not follow the general pattern are
\begin{equation*}
	\nabla_{\bm{\psi}_1}\ts\nabla_{\bm{\psi}_2}{\ell}  = 
	\begin{cases}
		{\bf M}\ts_{\bm \beta}\diag{{\boldsymbol \ell}^{\eta \tilde{s}}}{\bf M}_{\bm \alpha}+({\bf M}^{1}_{\bm \beta})\ts\diag{{\boldsymbol \ell}^{\eta}}{\bf M}_{\bm \alpha} & {\bm \psi}_1 = \bm{\alpha},\, {\bm \psi}_2 = \bm{\beta}\\
		{\bf M}_{\bm a}\ts\diag{{\boldsymbol \ell}^{\tilde{s}\tilde{s}}}{\bf M}_{\bm a}+\sum_{i=1}^n \ell_{i}^{\tilde{s}}\nabla_{\bm{a}}\ts\nabla_{\bm{a}}{\tilde s}_{i} & {\bm \psi}_1 = \bm{a},\, {\bm \psi}_2 = \bm{a}\\
		({\bf M}_{\bm b})\ts\diag{{\boldsymbol \ell}^{\eta \tilde{s}}}{\bf M}_{\bm a}+({\bf M}_{\bm b}^{1})\ts\diag{{\boldsymbol \ell}^{\eta}}{\bf M}_{\bm a} & {\bm \psi}_1 = \bm{a},\, {\bm \psi}_2 = \bm{b},
	\end{cases}
\end{equation*}
Such non-standard blocks involve derivatives w.r.t the parameters of a single nested effect. In contrast, blocks corresponding to pairs of parameter vectors belonging to distinct nested effects follow the general pattern \eqref{eq:gen_Hess_pattern}.

The general expressions for the Hessian blocks provided here apply to multi-parameter GAMs containing any combination of standard and nested effects. The computational cost of any block is 
$O(np_1p_2)$, where $p_1 = \text{dim}(\bm \psi_1)$ and $p_2 = \text{dim}(\bm \psi_2)$. However, the cost of computing ${\bf M}_{\bm a}$ and $\nabla_{\bm{a}}\ts\nabla_{\bm{a}}{\tilde s}_{i}$ is transformation-specific, while that of computing the derivatives of $\ell_i$ w.r.t. $\theta_{ij}$, $\eta_{ij}$ or $\tilde{s}_i$ depends on the response distribution.





\subsubsection{Derivatives of the Hessian blocks w.r.t. $\rho$} \label{sec:D_hess_D_rho}
Here we explain how to compute the derivative of the Hessian blocks, evaluated at the MAP estimate $\hat{\bm \zeta}$, w.r.t. a log-smoothing parameter $\rho = \log\lambda$. Let $L$ be the total number of $\bm{\gamma}$, $\bm{\alpha}$, $\bm{\beta}$,
$\bm{a}$ and $\bm{b}$ vectors contained in the model. Assume, for simplicity, that each vector has $p$ elements. Then  
\begin{equation*}	\pdif{\nabla_{\bm{\psi}_{j}}\ts\nabla_{\bm{\psi}_{k}}\ell}{\rho}
	= \sum_{l=1}^{L} \sum_{v=1}^{p}
	  \pdif{\nabla_{\bm{\psi}_{j}}\ts\nabla_{\bm{\psi}_{k}}\ell}{\hat{\psi}_{lv}}
	  \dif{\hat{\psi}_{lv}}{\rho}=\sum_{l=1}^{L}\bm{\Gamma}_{\bm{\psi}_l}^{\bm{\psi}_j\bm{\psi}_k},
\end{equation*}
where $d\hat{\bm \psi}_{l}/d{\rho}$ is the derivative of the $\hat{\bm \psi}_{l}$ sub-vector of $\hat{\bm \zeta}$, which can be computed by implicit differentiation, as in \cite{wood_smoothing_2016}. The terms $\bm{\Gamma}_{\bm{\psi}_l}^{\bm{\psi}_j\bm{\psi}_k}$ indicates the partial effect of $\rho$ on the $j$-$k$ block of the Hessian \emph{via} the $l$-th vector of coefficients. While na{\"i}ve evaluation of these derivatives would lead to a computational cost of $O(np^3)$, we reduce this to $O(np^2)$ for most terms, by following an approach similar to that of \cite{wood_smoothing_2016}.

Indicate
$\bm{\Gamma}_{\bm{\psi}_l}^{\bm{\psi}_j\bm{\psi}_k}$ simply with $\bm{\Gamma}_{\bm{\psi}_3}^{\bm{\psi}_1\bm{\psi}_2}$. Most such terms follow the general pattern 
\begin{equation}\label{eq:general_block}
	\bm{\Gamma}_{\bm{\psi}_3}^{\bm{\psi}_1\bm{\psi}_2}={\bf M}_{\bm{\psi}_{2}}\ts{\bf V}_{\bm{\psi}_{3}}^{\bm{\psi}_{1}\bm{\psi}_{2}}{\bf M}_{\bm{\psi}_{1}},	
\end{equation}
where ${\bf M}_{\bm{\psi}}$ is defined as before, while ${\bf V}_{\bm{\psi}_{3}}^{\bm{\psi}_{1}\bm{\psi}_{2}}$ is a diagonal matrix with non-zero elements
\begin{equation*}
	\left({\bf V}_{\bm{\psi}_{3}}^{\bm{\psi}_{1}\bm{\psi}_{2}}\right)_{ii}=\ell_{i}^{\nu(\bm{\psi}_{1}, \bm{\psi}_{2}, \bm{\psi}_{3})}\nu_{\bm{\psi}_{3}}(\bm{\psi}_{3})_{i}.
\end{equation*}
Here $\nu_{\bm{\psi}}(\bm{\psi})_{i}$ denotes the $i$-th element of the vector 
\begin{equation*}
	\nu_{\bm{\psi}}(\bm{\psi})
	= \dif{_{\bm \psi} \nu(\bm{\psi})}{_{\bm \psi}\rho}
	= {\bf M}_{\bm{\psi}} \dif{\hat{\bm \psi}}{\rho},
\end{equation*}
where $\dif{_{\bm \psi} \nu(\bm{\psi})}{_{\bm \psi}\rho} $ is the derivative of $\nu(\bm{\psi})$ w.r.t. $\rho$ via the $\hat{\bm \psi}$ sub-vector of $\hat{\bm \zeta}$. 
Evaluating \eqref{eq:general_block} has an $O(np^2)$ computational cost. However, some terms do not follow the pattern described above. Among these, only those of type $\bm{\Gamma}_{\bm{a}}^{\bm{a}\bm{a}}$ require $O(np^3)$ operations. See SM~\ref{app:exceptions_dH} for more details.

\subsection{Inference and model selection}

Adopting the Bayesian view of the smoothing process allows us to quantify parameter uncertainty. In particular, following \cite{wood_smoothing_2016}, we employ standard asymptotic techniques to approximate the posterior distribution of ${\bm \zeta}$, $p({\bm\zeta}\,|\,{\bf y}, {\bm \lambda}) $, with a Gaussian centred at the MAP estimate $\hat{\bm \zeta}$ and with covariance ${\bf V}_{\bm \zeta} =(\hat{\bm{\mathcal{I}}} + {\bf S}_{\bm \lambda})^{-1}$, where $\hat{\bm{\mathcal{I}}}$ is the Hessian of the negative log-likelihood. That is, ${\bm \zeta}\,|\, {\bf y}, {\bm \lambda}\sim \mathcal{N}(\hat{\bm \zeta},{\bf V}_{\bm \zeta}) $. This approximation treats the smoothing parameters as fixed at the LAML maximiser and therefore ignores their uncertainty. In principle, an approximation to the unconditional posterior $p({\bm \zeta}\,|\,{\bf y})$ could be obtained by applying a Gaussian approximation to $p({\bm \lambda}\,|\, {\bf y})$ and propagating smoothing parameter uncertainty forward, as done by \cite{wood_smoothing_2016}. However, doing so would require the Hessian of $\tilde{\mathcal{V}}({\bm\lambda})$ w.r.t. ${\bm \lambda}$, which involves computing the fourth-order derivative of log-likelihood w.r.t. the elements of $\bm \eta$. Given the complexity of the derivative system presented above, we leave this extension to future work. 

The Bayesian posterior distribution of the nested effects can be derived by propagating the approximate posterior distribution of $\bm \zeta$ via the delta method.
In particular, the asymptotic posterior distribution of the nested effect vector ${\bf s} = s[\tilde{s}({\bf x}^{\tilde{S}})]$ can be approximated by a Gaussian distribution with mean equal to the estimated nested effect $\hat{\bf s} = \hat{s}[\hat{\tilde{s}}({\bf x}^{\tilde{S}})]$ and $n\times n$ covariance matrix ${\bf V}_{\bf s} = \nabla_{\bm a \bm b}\ts \hat{\bf s} \left[{\bf V}_{\bm \zeta}\right]_{\bm a \bm b}\nabla_{\bm a \bm b} \hat{\bf s}$. Here $\left[{\bf V}_{\bm \zeta}\right]_{\bm a \bm b} $ is the block of ${\bf V}_{\bm \zeta}$ representing the posterior covariance matrix of $\bm{ a}$ and $\bm b$, while $\nabla_{\bm a \bm b}\ts \hat{\bf s} = [\operatorname{diag}({\bf M}^1_{\bm b}{\hat{\bm b}}) {\bf M}_{\bm a}, {\bf M}_{\bm b}]$ with $(\nabla_{\bm a \bm b}\ts \hat{\bf s})_{ij} = \partial {s}_i / \partial a_j|_{{\bm \zeta} = \hat{\bm \zeta}}$, if $j \leq d_{\bm a} = \text{dim}({\bm a})$, and $(\nabla_{\bm a \bm b}\ts \hat{\bf s})_{ij} = \partial {s}_i / \partial b_{j-d_{\bm a}}|_{{\bm \zeta} = \hat{\bm \zeta}}$, if $j > d_{\bm a}$. 

The use of improper smoothing priors invalidates Bayesian model selection via marginal likelihood methods. 
An alternative is to use the Akaike information criterion which, for penalised GAMs, can be defined as $\operatorname{AIC} = -2\ell(\hat{\bm{\zeta}}) + 2\hat{\tau}$, where 
$\hat{\tau} = \tr{\bf F}$ represents the effective degrees of freedom (e.d.f.) and ${\bf F} = {\bf V}_{\bm{\zeta}} \bm{\hat{\mathcal{I}}}$ \citep{wood_smoothing_2016}. 
This e.d.f. definition can be seen as a by-product of Wood's \citeyear[][]{wood_smoothing_2016} derivation of the AIC, which focused on estimating the out-of-sample predictive performance of penalised GAMs. However, here we demonstrate that, when $y_i = \mu_i + \varepsilon_i$ with $\varepsilon_i\sim \mathcal{N}(0, \sigma^2)$, it matches the more explicit e.d.f. definition of \cite{Efron01061986}, that is $\hat{\tau}_E = \sigma^{-1}\operatorname{cov}(\hat{\bm \mu}, {\bf y})$.

We start by applying Stein's lemma \citep{stein1981estimation} to the definition of $\hat{\tau}_E$, which leads to
\begin{equation}\label{eq:edf_stein}
\hat{\tau}_E = \sum_{i=1}^n \pdif{\hat{\mu}_i}{y_i} = \sum_{i=1}^n \pdif{\hat{\mu}_i}{\hat{\bm{\zeta}}\ts}\pdif{\hat{\bm{\zeta}}}{y_i}.
\end{equation}
Let $\ell = \sum_i \ell_i$ be the log-likelihood. By definition, $\hat{\bm{\zeta}}$ satisfies $$\left(\pdif{\ell}{\bm{\zeta}} - {\bf S}_{\bm{\lambda}}\bm{\zeta}\right)\Bigg|_{\bm{\zeta} = \hat{\bm{\zeta}}} = 0,$$ 
and, by differentiating both sides w.r.t. $y_i$, we find
$$
\pdif{\hat{\bm{\zeta}}}{y_i} = -\left(-\frac{\partial^2 \ell}{\partial\bm{\zeta}\ts \partial\bm{\zeta}}\Bigg|_{\bm{\zeta} = \hat{\bm{\zeta}}} + {\bf S}_{\bm{\lambda}}\right)^{-1} \pddif{\ell_i}{{\mu}_i}{y_i}\Bigg|_{\bm{\zeta} = \hat{\bm{\zeta}}}\pdif{\hat{\mu}_i}{\hat{\bm{\zeta}}}.$$ 
Plugging this into the definition of $\hat{\tau}_E$ leads to 
\begin{equation}\label{eq:edf1}
    \begin{aligned}
        \hat{\tau}_E &= \sum_{i=1}^n \pdif{\hat{\mu}_i}{\hat{\bm{\zeta}}\ts}
        \left(-\frac{\partial^2 \ell}{\partial\bm{\zeta}\ts \partial\bm{\zeta}}\Bigg|_{\bm{\zeta} = \hat{\bm{\zeta}}}+
        {\bf S}_{\bm{\lambda}}\right)^{-1} 
        \pddif{\ell_i}{{\mu}_i}{y_i}\Bigg|_{\bm{\zeta} = \hat{\bm{\zeta}}}\pdif{\hat{\mu}_i}{\hat{\bm{\zeta}}}\\
        & = \frac{1}{\sigma^2}\operatorname{tr}\left[\left(\hat{\bm{\mathcal{I}}}+{\bf S}_{\bm{\lambda}}\right)^{-1} 
            \sum_{i=1}^n\pdif{\hat{\mu}_i}{\hat{\bm{\zeta}}}\pdif{\hat{\mu}_i}{\hat{\bm{\zeta}}\ts}\right] = \tr{\bf F_{\hat{\tau}}}.
    \end{aligned}
\end{equation}
Under the Gaussian assumption $\sigma^{-2} \sum_{i=1}^n {\partial\hat{\mu}_i}/{\partial\hat{\bm{\zeta}}}( {\partial\hat{\mu}_i}/{\partial\hat{\bm{\zeta}}\ts}) = \hat{\bm{\mathcal{I}}}$, thus $\hat{\tau}_E$ is the trace of ${\bf F}_{\hat{\tau}} = (\hat{\bm{\mathcal{I}}} + {\bf S}_{\bm\lambda})^{-1}\hat{\bm{\mathcal{I}}} = {\bf V}_{\bm{\zeta}} \bm{\hat{\mathcal{I}}} = {\bf F}$. Therefore, the explicit e.d.f. definition of \cite{Efron01061986} matches that of \cite{wood_smoothing_2016}, which was based on a different line of reasoning.  

\section{Applications}\label{sec:applications}

Here we illustrate the effectiveness of the nested smooth effects in the context of two challenging applications. Recall that our fitting framework requires the basis functions underlying the nested smooth effects to be four times differentiable. We use B-spline bases of sixth degree (i.e., sextic), which fulfil this requirement and have derivatives that are readily computed via the \verb|splines| \verb|R| package. All nested effects are smoothed via second-order derivative penalties.  Unless stated otherwise, standard smooth effects are constructed using thin plate spline bases and regularised via second derivative penalties. 

\subsection{Electricity net-demand in Great Britain}\label{sec:application_netdemand}

Electricity net-demand is the demand minus embedded generation, measured at the interface between the high-voltage transmission grid and a distribution network. In Great Britain (GB), these interfaces are referred to as Grid Supply Points (GSP) and are organized into 14 regions, known as GSP groups. We are interested in forecasting the total net-demand in GB, that is, its sum across the GSP groups, one day ahead. Such forecasts are key inputs for many operations in the electricity industry, such as trading and production planning. We focus on the net-demand between 12:00 AM and 12:30 AM, from January 2014 to December 2018. As predictors, we use calendar information such as bank/school holidays, weekdays, and day of the year, and day-ahead weather forecasts produced by the operational ECMWF-HRES model. The raw weather predictions are available on a spatial grid, but have been reduced to 14 regional forecasts, following the approach of \cite{browell2021probabilistic}. The available covariates are listed in Table~\ref{tab:covariates net-demand}.

\begin{table}[t]
\footnotesize
\renewcommand{\arraystretch}{1.29}
\begin{center}  
\begin{tabular}{ l @{\hskip 0.08cm}  p{6.60cm}   | l @{\hskip 0.08cm} p{6.60cm} } 
\multicolumn{2}{l|}{\textbf{General covariates}} & \multicolumn{2}{l}{\textbf{Covariates derived from weather forecasts}} \\
\toprule
$\text{t}_i$ & time since the 1st January 2014 & $\text{rain}_{ij}$ & mean precipitation ($mm \ h^{-1}$) \\ 
$\text{dow}^+_{i}$ & day of the week factor with additional factor levels accounting for public holidays & $\text{temp}_{ij}$ & temperature (K) at cell with highest regional population density \\
$\text{wcap}_i$ & GB embedded wind generation capacity (MW) & $\text{irr}_{ij}$ & mean solar irradiance (W$\ m^{-2}$) times embedded solar generation capacity (MW) \\   
$\text{doy}_{i}$ & day of the year ($\in \{1, \dots, 366\}$) & $\text{wsp}^{100}_{ij}$ & mean wind speed at $100$ metres ($m\ s^{-1}$) \\  
$\text{shol}_{i}$ & school holidays, three levels factor to distinguish Christmas from other holidays &  & \\
$y_{i}^{h}$ & net-demand at $h$ hours lag &  & \\
$\text{st}_{i}$ & factor denoting the presence of a storm in GB &  & \\
\bottomrule
\end{tabular}  
\caption{Variables used to model net-demand in Great Britain. The subscript $i$ indicates the $i$-th observation, and $j$ denotes the $j$-th GSP group. }\label{tab:covariates net-demand}
\end{center} 
\end{table}

Let $y_i, \ i=1,\ldots, n$ be the net-demand, and assume that $y_i\sim\mathcal{N}(\mu_i, \sigma_i^2)$ with mean and variance modelled via
\begin{equation}
    \label{eq:netdemand}
    \begin{aligned}
        \mu_i &=g_1(\text{t}_i) + g_2( \text{dow}^+_i) + g_3(\text{shol}_i)
        + f^{30}_1(\text{doy}_i) + f_2^{10}(\widebar{\text{irr}}_i) + f_3^{10}\left(\widebar{\text{rain}}_i^{\frac{1}{2}}\right)\\ 
            &\qquad  +s_1^{10}(\text{wcap}_i\cdot\textbf{wsp}_i\ts{\bm a}^{w}) + s_2^{10}\left[(\mathbf{y}_{i}^L)\ts{\bm a}^{y}\right]  \\
            &\qquad +s_3^{15}\left[\tilde{s}_{3}^{\text{exp}}(\widebar{\text{temp}}_i)\right] + s_4^{10}\left[\tilde{s}_{4}^{\text{exp}}(\widebar{\text{temp}}_i)\right]\\
        \log{\sigma_i}^2 &= g_4(\text{st}_i) +  g_5(\text{shol}_i) + f_4^{10}(\text{doy}_i),
    \end{aligned}
\end{equation}
where $g_1$ to $g_5$ are parametric (linear) effects, $f_1$ to $f_4$ are standard smooth effects and $s_1$ to $s_4$ are smooth effects with nested transformations. The superscripts of the $f$'s and $s$'s indicate the number of spline basis functions used. The spline bases and penalties are those described at the beginning of this section except for  $f^{30}_1(\text{doy}_i)$, which uses a B-spline basis with an adaptive P-spline penalty \citep{eilers1996flexible} designed to allow the smoothness to vary with the covariate \citep[see Section 5.3.5 of][for details]{Wood2017}. 

The model includes four nested transformations. The first is a linear combination, with single index vector ${\bm a}^{y}$, of past half-hourly net-demand values $\mathbf{y}_{i}^L$, with lags ranging from 12 to 34 hours before $y_i$ is observed. 
Lags between 34.5 and 36.5 hours are excluded because the corresponding demand dynamics are unstable, owing to the persistent effects of transitions between British Summer Time and Greenwich Mean Time. The elements of ${\bm a}^{y}$ are regularised via the second-order difference penalty $\sum_l ({a}^y_{l+1} - 2{a}^y_{l} + {a}^y_{l-1})^2$, which encourages them to vary linearly with their index. A further linear combination, with coefficient vector ${\bm a}^{w}$, is used to reduce the regional wind speed forecasts to a single index, which is then scaled by GB wind generation capacity. We expect its elements to be proportional, in absolute value, to the embedded wind production capacity in each region. 

The remaining weather forecasts have been summarised by taking their sample mean across the GSP groups, resulting in the scalar valued covariates $\widebar{\text{irr}}_i$, $\widebar{\text{rain}}_i$ and $\widebar{\text{temp}}_i$. To justify this choice, note that $\text{irr}_{ij}$ is already scaled by the installed solar production capacity in the $j$-th region, so estimating regional weights, as we did for wind speed, seems unnecessary. During model development, we experimented with single index effects for precipitation and temperature, but the estimated weights did not vary significantly across the regions. Hence, we opted for standard smooth effects of their mean values across GB.   

\begin{figure}
    \includegraphics[width=\textwidth]{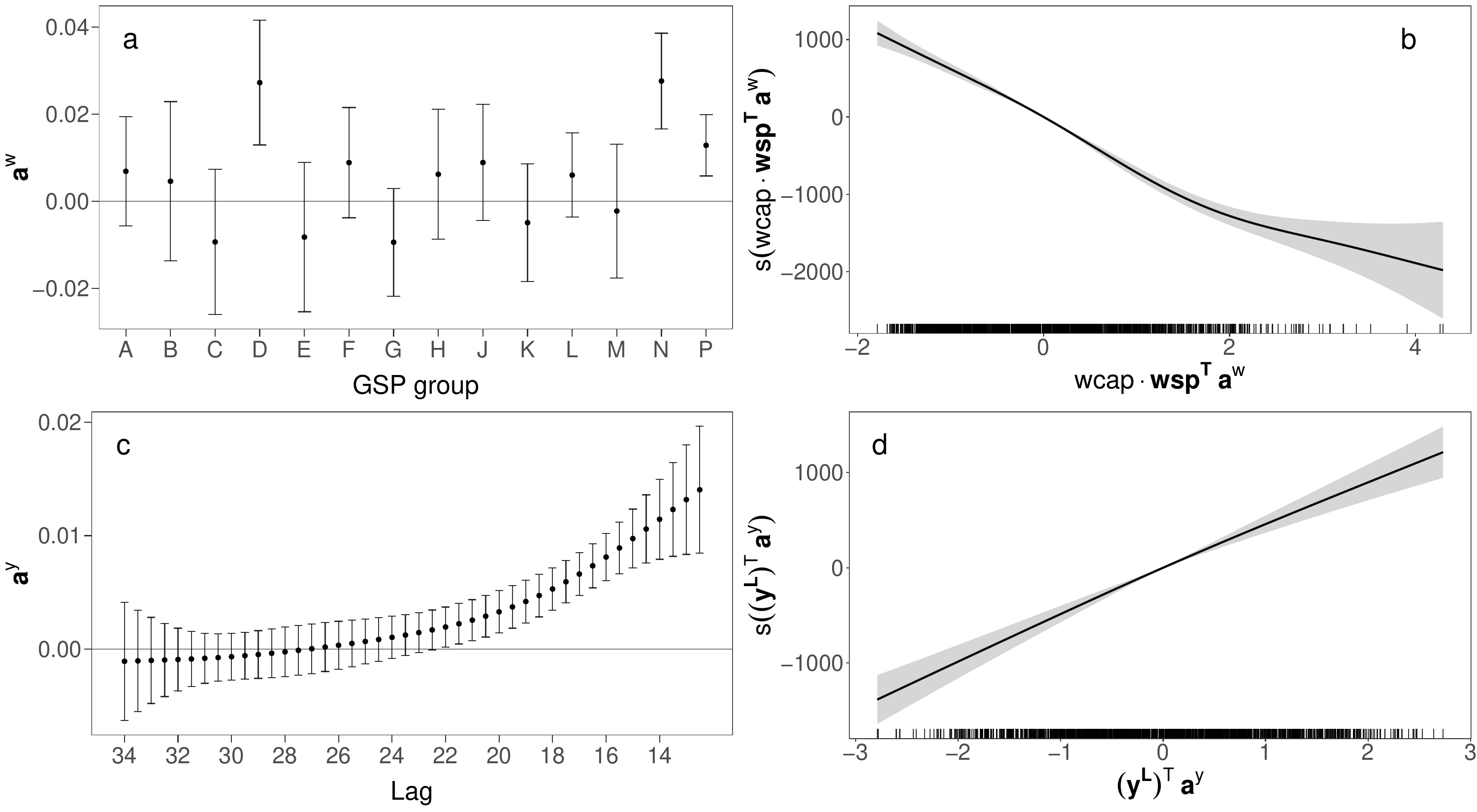}
    \caption{Inner single index coefficients and outer smooth effect of wind speed (a-b) and of net-demand lags (c-d).} \label{fig:net-demand effects}
\end{figure}

The third and fourth nested effects in (\ref{eq:netdemand}) are based on exponential smooth effects of average GB temperature. Such effects are meant to capture thermal inertia, which might evolve at an intra-day scale. However, we are forecasting only one value of $y_i$ per day, hence the effects must handle different temporal resolutions. In particular, exponential smoothing is performed on a half-hourly resolution, that is $\tilde{s}_{3-4}^{\text{exp}}(\widebar{\text{temp}}_i)$  use the $48\times i$ temperature forecasts that are available before the $i$-th net-demand value is observed, but only the smoothed temperature value corresponding to 12 AM is used to predict $y_i$. The irregular time gaps induced by missing temperature values are handled by letting the exponential smoothing rate vary with $\tilde{\mathbf{x}}_i\ts {\bm a}^{\text{exp}}$, where $\tilde{\mathbf{x}}_i = [1, \Delta h_i - 1]$ and $\Delta h_i$ is the number of hours between $\widebar{\text{temp}}_i$ and the preceding temperature forecast. 

We fitted model (\ref{eq:netdemand}) on data from 2014 to 2017, leaving 2018 for testing. Figure~\ref{fig:net-demand effects} shows some of the estimated effects and inner transformations. The top row shows the linear combination coefficients of wind speed (a) and the corresponding smooth effect (b). As expected, the single index elements that are significantly different from zero correspond to GSP groups with substantial wind production capacity, namely North Wales and Merseyside (D), South and Central Scotland (N), and North Scotland (P). Considering that the smooth effect of wind speed is monotonically decreasing, this suggests that, as wind speed increases, net-demand decreases due to the growth in embedded wind production. In contrast, the coefficient for London (C) is negative, hence net-demand increases with wind speed in this region. Even though the coefficient is not significantly different from zero and should not be over-interpreted, its sign is more likely to be related to the cooling effect of wind speed, rather than to wind production, which is negligible in this highly urbanised area. 

The bottom row of Figure~\ref{fig:net-demand effects} shows the single index coefficients of lagged net-demand (c) and the corresponding effect (d). In accordance with intuition, the overall effect of lagged net-demand is positive, as all the non-zero coefficients are positive and the smooth effect is monotonically increasing. Further, the coefficients decrease monotonically as the lag increases, meaning that recent net-demand values are more informative for predicting future net-demand. Note that this was not obvious a priori, because net-demand behaves differently depending on the hour of the day. In particular, here the most recent lag, which corresponds to 11:30 PM, has been assigned the highest weight when predicting net-demand at 12 AM. 

The exponentially smoothed temperatures and their effect on net-demand are shown in Figure \ref{fig:first_examples}a-b. The two effects converged on different smoothing regimes, corresponding to exponential parameters approximately equal to $0.19$ (red) and 0.996 (blue). The red smoothed temperature trajectory corresponds to a low thermal inertia regime. It affects net-demand via a U-shaped effect, which arguably captures both heating and cooling. Instead, the blue temperature trajectory, is characterised by higher inertia, and it affects net-demand only via cooling. This could be due to the fact that heat-waves are usually of short duration in GB, hence they are effectively filtered out by the smoother temperature trajectory. 

To validate the model, we compare its predictive performance with that of four alternative models on the 2018 test data. All such models assume a Gaussian response and model $\log{\sigma_i}^2$ as in \eqref{eq:netdemand}. Their mean parameter is controlled by the effects appearing on the first row of \eqref{eq:netdemand}, that is from $g_1$ up to $f_3$ (omitted below), but the remaining effects have been replaced with the following
\begin{align}
        \mu_i^1 &= \cdots + f_4^{15}(\widebar{\text{temp}}_i) + f_5^{10}(\widebar{\text{wind}}_i)+f_6^{10}(y_{i}^{24}),\label{eq:netdemand_1}\tag{M1} \\
        \mu_i^2 &= \cdots + f_4^{15}(\widebar{\text{temp}}_i) + f_5^{10}(\widebar{\text{wind}}_i)+f_6^{10}(\mathbf{y}_{i}^L), \label{eq:netdemand_2}\tag{M2}\\
        \mu_i^3 &= \cdots + f_4^{15}(\widebar{\text{temp}}_i) + f_5^{10}(\widebar{\text{wind}}_i)+f_6^{10}(\mathbf{y}_{i}^L)+ f_8^{10}(\widebar{\text{temp}}^{95}_i), \label{eq:netdemand_3}\tag{M3}\\
        \mu_i^4 &= \cdots + f_4^{15}(\widebar{\text{temp}}_i) + f_5^{10}(\widebar{\text{wind}}_i)+f_6^{10}(\mathbf{y}_{i}^L)+ f_8^{10}(\widebar{\text{temp}}^{95}_i)\label{eq:netdemand_4}\tag{M4}\\
        & \qquad+ s_5^{10}(\text{wcap}_i\cdot\textbf{wsp}_i\ts{\bm a}^{wsp}). \notag
\end{align}
Here $\widebar{\text{wind}}_i$ is the sample mean of wind speeds across the GSP groups and $f_6^{10}(\mathbf{y}_{i}^L)$ is a distributed lag effect. The latter is a type of functional smooth effect obtained by building a bivariate tensor-product smooth effect of lagged net-demand and the corresponding lag, which is then integrated over the lag. It is an obvious alternative to the single index effect $s_2[(\mathbf{y}_{i}^L)\ts\bm{a}^y]$ in \eqref{eq:netdemand}. A simpler alternative is provided by M1, which includes the effect of net-demand at a 24-hour lag only. Variable $\widebar{\text{temp}}^{95}_i$ is an exponential smooth of the average GB temperature with smoothing coefficient fixed a priori to 0.95, which is a commonly used value in aggregate demand forecasting applications \citep[see, e.g.,][]{GAILLARD20161038}. The last effect in M4 is a single index effect of wind speed, where the index coefficients have been estimated by maximising the profile LAML in an outer iteration, which requires refitting the model multiple times, rather than via the methods proposed here.


\begin{table}\centering
\begin{tabular}[t]{l|cccccc}
\toprule
  & \textbf{Log-score} & \textbf{CRPS} & \textbf{RMSE} & \textbf{MAE} & \textbf{AIC} & \textbf{Time (min)}\\
    \midrule
M1 & 2672 & 134893 & 726 & 563 & 21694 & \underline{0.5}\\
M2 & 2660 & 128597 & 693 & 538 & 21567 & 0.7\\
M3 & 2648 & 122555 & 658 & 511 & 21504 & 0.8\\
M4 & 2623 & 112791 & 617 & 472 & 21304 & 1160.0\\
Model \eqref{eq:netdemand} & \underline{2621} & \underline{111634} & \underline{614} & \underline{463} & \underline{21299} & 10.0\\
    \bottomrule
\end{tabular}
\caption{Comparison of out-of-sample performance in the electricity net-demand application, using nested versus standard effects. The best score in each column is underlined.}\label{tab:elect perf}
\end{table}

The first three columns of Table~\ref{tab:elect perf} report the performance of each model in terms of out-of-sample negative log-likelihood (log-score), marginal continuous ranked probability score (CRPS) and root mean squared error (RMSE). The table shows a clear improvement in performance as model complexity increases from M1 to M4, with model \eqref{eq:netdemand} leading to the lowest losses. The AIC of each model in the fourth column are in accordance with the out-of-sample losses. The last column reports the fitting times, and show that the proposed model is around twenty times slower than standard GAMs with no nested effects (M1-M3). 




\subsection{Modelling house prices in London} \label{sec:application_house}
Here we focus on modelling how house prices are affected by local socio-economical factors, as well as by characteristics of the property being sold, while accounting for spatial autocorrelation. We consider publicly available price paid data in London during 2022, provided by HM Land Registry. In addition to sales prices and the corresponding postcodes, the data contains several categorical variables providing information on each property,  specifically: property type, $\text{type}_i$ (detached, semi-detached, terraced, flats/maisonettes or other), the age of the property, $\text{new}_i$ (newly built or established residential building), type of price paid, $\text{cat}_i$ (standard or additional), and type of ownership, $\text{free}_i$ (freehold or leasehold).  
We integrate the data with the Index of Multiple Deprivation (IMD), which is a composite measure relative deprivation. The index is provided on small areas, comprising between 400 and 1200 households, called  Lower layer Super Output Areas (LSOAs). Further, we compute the Euclidean distance, $d_i$, of each property from the nearest underground station. See SM~\ref{sec:source of data} for more details on the data.

The full data set comprises 69201 sales, reduced to 67686 after excluding properties sold for less than £100,000 or more than £10 million. We exclude extremely low or high transactions prices because they might be heavily affected by exceptional circumstances, hence entirely unrelated to factors of interest, such as the distance from the tube (the highest price in the full data is £429 million). Further, postcodes are highly localised in the UK, for example a large block of flats can be attributed a unique postcode, leading to highly correlated postcode-level prices. Hence, we average prices at postcode level, thus obtaining a data set of 34537 observed prices.

\begin{figure}[t]
    \begin{subfigure}{\textwidth}\centering
        \includegraphics[width =0.42\textwidth]{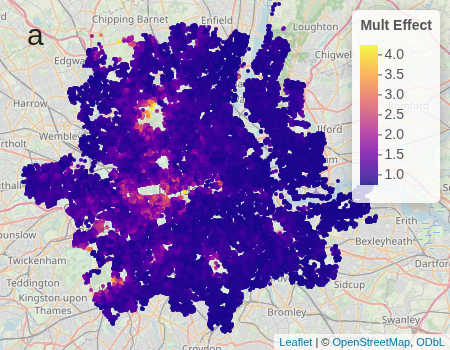}\quad
        \includegraphics[width =0.42\textwidth]{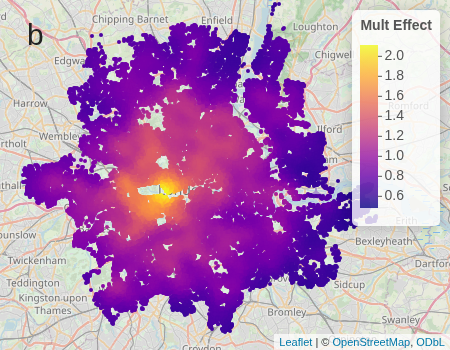}\\
        \includegraphics[width =0.42\textwidth]{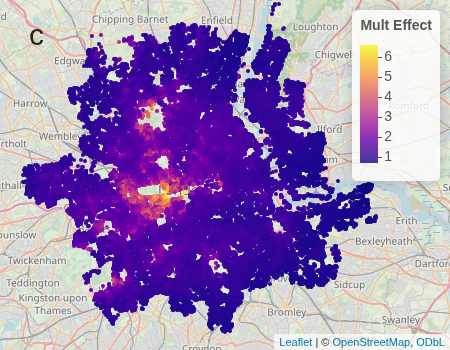}\quad
        \includegraphics[width =0.42\textwidth]{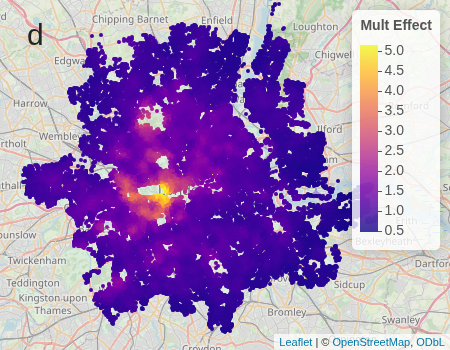}
    \end{subfigure}
    \caption{The effect of neighbouring prices, $s^{10}\left[\tilde{s}^{\text{mgks}}(\mathbf{x}_{i})\right]$ (a), the spatial effect, $f_3^{400}(\text{lon}_i, \text{lat}_i)$ (b) and their joint effect on expected prices (c), estimated under model (\ref{eq:mod_house}). The estimated spatial effect, $f_3^{2000}(\text{lon}_i, \text{lat}_i)$ (d), under a standard model that does not include the spatial autoregressive term. The effects have been exponentiated to obtain multiplicative effects on the original, rather than logarithmic, price scale.} \label{fig:maps house data}
\end{figure}

Let $y_i$ be the logarithm of the averaged price in postcode $i$. We model its conditional distribution using a GAMLSS model based on the sinh-arcsinh distribution of \cite{jones2009sinh}, which has four parameters $\mu$, $\sigma>0$, $\epsilon$, and $\delta>0$, controlling location, scale, skewness, and kurtosis. The model allows for asymmetry to either side, depending on the sign of $\epsilon$, and can have lighter ($\delta>1$) or heavier ($0<\delta<1$) tails than a Gaussian distribution. We model $\epsilon$ and $\delta$ only via intercepts, while the location and scale parameters are controlled by
\begin{equation}\label{eq:mod_house}
    \begin{aligned}
        \mu_i &= g_1(\text{type}_i) + g_2(\text{new}_i) + g_3(\text{free}_i) + g_4(\text{cat}_i) + f_1^{10}(\text{imd}_i)\\
        &\qquad  + f_2^{10}(\text{d}_i^{\frac{1}{2}}) +f_{3}^{400}(\text{lon}_i,\text{lat}_i)+ s^{10}\left[\tilde{s}^{\text{mgks}}(\mathbf{x}_{i})\right]\\
        \log(\sigma^2_i)&=g_5(\text{type}_i) + g_6(\text{cat}_i) + f_4^{10}(\text{imd}_i).
    \end{aligned}
\end{equation}
As in the net-demand application, the $g$'s indicate parametric effects and the $f$'s standard smooth effects. The latter are based on the spline bases and penalties described at the beginning of this section, including $f_3$, which is an isotropic bivariate spatial effect.

The last smooth effect used to model $\mu_i$ in (\ref{eq:mod_house}) contains a nested bivariate Gaussian kernel smoothing transformation of neighbouring house log-prices. In particular
\begin{equation}\label{eq:mgks_house}
    \tilde{s}^{\text{mgks}}({\bf x}_i) = \frac{\sum_{j\in \mathcal{N}_i}K_{{a}}({\bf x}_{i}, {{\bf x}}_{j})y_{j}}{\sum_{q\in \mathcal{N}_i}K_{{a}}({\bf x}_{i},{{\bf x}}_{q})}, 
\end{equation}
where ${\bf x}_i = \{\text{lon}_i, \text{lat}_i\}$, $\mathcal{N}_i \not\ni i$ is the set of the $L$ postcodes that are closest to $i$ in Euclidean distance, $K_i$ is the bivariate Gaussian p.d.f with diagonal, isotropic covariance matrix, parametrised by $a$. The purpose of this smooth is to capture the effect of neighbouring prices via a spatial autoregressive model, where the rate of decay of the spatial weights is controlled by $a$. Here we use $L = 25$ neighbours to reduce the computational cost, but the results below are unchanged by setting $L = 100$. 

To quantify the importance of including the spatial autoregressive component in (\ref{eq:mod_house}), we consider a sequence of models that do not include such an effect. In particular, we remove the autoregressive term and we fit models that only use $f_{3}^{k}(\text{lon}_i,\text{lat}_i)$, with different basis dimension $k$, to model the spatial variation of house log-prices. We test the model on a test set, generated by random splitting the postcodes between a train ($75\%$ of the data) and a testing set ($25\%$).

\begin{figure}
    \includegraphics[width = \linewidth]{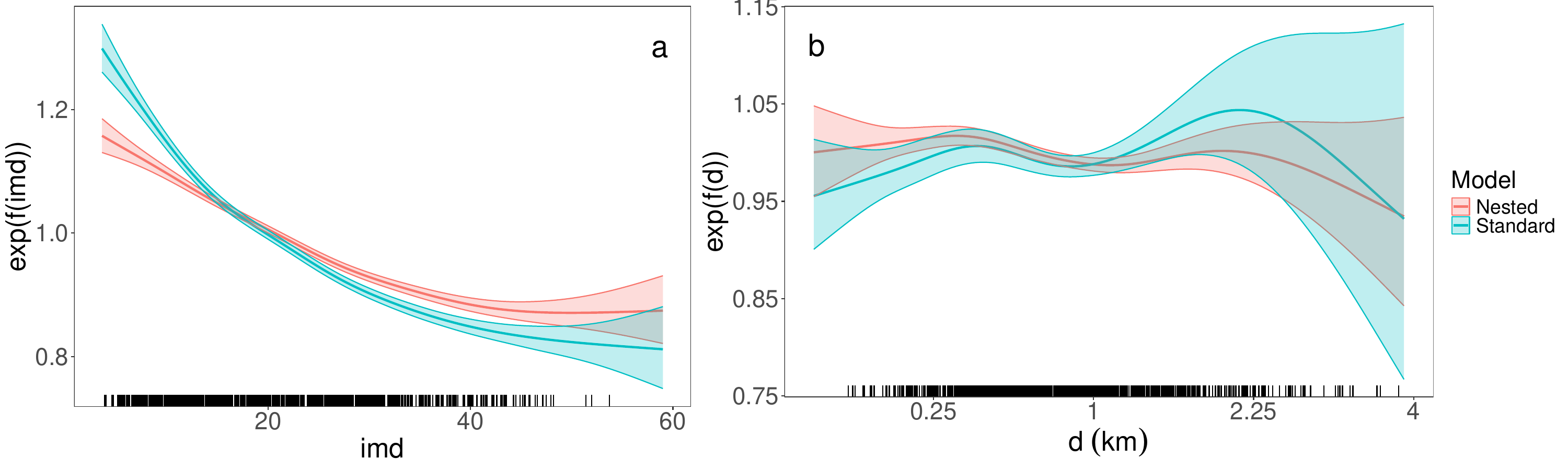}
    \caption{Multiplicative effects of IMD (a) and distance to the nearest tube station (b), estimated by model  \eqref{eq:mod_house} and a model using $k = 2000$ basis functions for the standard spatial effect, but no autoregressive effect. The $x$-axis in (b) is on a square root scale.
    } \label{fig:house prices effects}
\end{figure}

The results on the same loss functions considered in Section \ref{sec:application_netdemand} are shown in Table \ref{tab:house perf}. The second column reports the number of basis functions, $k$, used for the isotropic spatial effect in each model. It is interesting to note that achieving a predictive performance similar to that of the autoregressive models requires a highly parametrised spatial effect ($k = 2000$), which results in a computational time that is nearly twenty times longer.

\begin{table}[t]
\centering
\begin{tabular}{cc|cccccc}
  & \textbf{Knots}&\textbf{Log-score} & \textbf{CRPS} & \textbf{RMSE} & \textbf{MAE} & \textbf{AIC} & \textbf{Time (min)}\\ 
  \midrule
standard & 400 & 3899 & 1949 & 0.437 & 0.304 & 23029 & \underline{40}\\
standard & 1000 & 3807 & 1925 & 0.433 & 0.301 & 22105 & 194\\
standard & 1600 & 3777 & 1915 & 0.432 & 0.299 & 21779 & 444\\
standard & 2000 & 3771 & 1915 & 0.431 & 0.299 & 21715 & 1533\\
nested & \underline{400} & \underline{3690} & \underline{1895} & \underline{0.427} & \underline{0.297} & \underline{21653} & 84\\ \bottomrule
\end{tabular}
\caption{Comparison of out-of-sample performance in the house prices modelling application, using nested versus standard spatial effects, with varying basis dimensions. The best score in each column is underlined.}\label{tab:house perf}
\end{table}

Figure \ref{fig:maps house data} shows that the nested model (a-c) is able to separate the effect of the, highly localised, neighbouring prices shown in Figure \ref{fig:first_examples}d-e, from the smoother effect of space itself. Instead, in the standard model (d), short- and long-range price dynamics are confounded within a single spatial effect, which struggles to capture two different spatial resolutions.

Figure \ref{fig:first_examples}c shows the multiplicative effect of the local price estimated by $\ref{eq:mgks_house}$ on the expected price. The $x$-axis has been adjusted to represent real, rather than log-prices. The effect is centred, hence equal to one, when the local price is approximately £$740K$, the median local price being around £$680K$. Interestingly, the slope of the effect increases with the neighbouring prices, suggesting that the autoregressive effect is stronger on the upper end of the price range. Stronger spatial autocorrelation in affluent areas might be attributable to local characteristics not captured by the model, such as proximity to luxury developments, as well as by high-end prices being driven more by local reference prices and comparable sales than by fundamentals such as construction costs or rental values.

The effect of IMD and distance from the tube, estimated under model (\ref{eq:mgks_house}) and the standard model with $k = 2000$, are shown in Figure~\ref{fig:house prices effects}.  As expected, prices are inversely proportional to IMD, with a slightly stronger effect under the standard model. The effect of distance from the tube is weaker, with model (\ref{eq:mgks_house}) estimating a slight decrease in prices with distance. Interestingly, the effect flattens beyond one kilometre, which is coherent with recent studies \citep{bbc2025unknown}.


\section{Conclusion} \label{sec:conclusion}

This work introduces a novel framework for incorporating complex covariate transformations into multi-parameter GAMs, with particular emphasis on transformations that provide automatic feature-engineering capabilities. By extending the implicit differentiation methods of \cite{wood_smoothing_2016} to accommodate the structure of nested effects, we enable the joint estimation of regression coefficients and transformation parameters. Doing so removes the need for a separate data preprocessing step, as well as the costly re-fitting of the entire model to determine transformation coefficients. Additionally, it enables joint uncertainty quantification of all parameters through approximate Bayesian methods.

Motivated by applications in electricity net-demand forecasting and house-price modelling, we focus on three specific types of interpretable transformations designed to handle complex covariates, such as time-series and spatial data. However, the modular fitting and inferential framework developed here is sufficiently general to incorporate any scalar-valued function $\tilde{s}$ that is sufficiently differentiable w.r.t. its parameters, thus providing several directions for future research. For example, it would be interesting to consider  single index effects where the inner coefficients are constrained, as done by \cite{masselot_2022}. Doing so would enhance the interpretability of the results, especially when the linear transformation is viewed as an index or a weighted average. 



Finally, it is worth noting that the proposed framework would, in principle, permit the inclusion of composite transformations obtained by nesting several basic transformations. Incorporating such transformations into multi-parameter GAMs would bring them closer to deep learning models with embedded feature engineering capabilities, while still preserving the interpretability associated with an additive structure. This interpretability is crucial in high-stakes applications, such as net-demand forecasting for power-production planning, where understanding the rationale behind a model is essential. 

\subsection*{Data Availability Statement}\label{data-availability-statement}
The electricity net-demand data are available on Zenodo at \url{https://doi.org/10.5281/zenodo.5031704}. Section \ref{sec:source of data} lists the data sources for the house prices.
\subsection*{Disclosure statement}
The authors have no conflicts of interest to declare
\if1\blind
{
\subsection*{Acknowledgments}
Claudia Collarin's work has been partially funded by PON ``Research and Innovation'' 2014\,--\,2020 Action IV.5 ``PhDs on Green issues.''\,--\,Ministerial Decree 1061/2021. Matteo Fasiolo's work has been partially funded by EDF R\&D. This work contains HM Land Registry data \textsuperscript{\textcopyright} Crown copyright and database right 2021. This data is licensed under the Open Government Licence v3.0.
} \fi

%
%

\bibliographystyle{chicago}
\bibliography{biblio.bib}

\include{supplementary_mat}

\end{document}

%% file: supplementary_mat.tex
\pagebreak
\clearpage
\makeatletter
\begin{center}
\textbf{\large Supplementary material to ``\papertitle''}\\
\vspace{2mm}
\if1\blind
{
Claudia Collarin, Matteo Fasiolo, Yannig Goude, Simon N. Wood
} \fi
\end{center}
\makeatother
\setcounter{section}{0}
\setcounter{equation}{0}
\setcounter{Theorem}{0}
\setcounter{figure}{0}
\setcounter{table}{0}
\setcounter{page}{1}
\makeatletter
\renewcommand{\theequation}{S\arabic{equation}}
\renewcommand{\theTheorem}{S\arabic{Theorem}}
\renewcommand{\thefigure}{S\arabic{figure}}
\renewcommand{\thesection}{S\arabic{section}}
\renewcommand{\bibnumfmt}[1]{[S#1]}
\renewcommand{\citenumfont}[1]{S#1}


\renewcommand{\theHequation}{S.\arabic{equation}}
\renewcommand{\theHTheorem}{S.\arabic{Theorem}}
\renewcommand{\theHfigure}{S.\arabic{figure}}
\renewcommand{\theHsection}{S.\arabic{section}}
\renewcommand{\theHtable}{S.\arabic{table}}
\makeatother

\spacingset{1}

\section{Practical details on the solution to the scaling issue}\label{app:scaling_penalty}
In Section~\ref{sec:scaling_issues}, we presented the scaling and knots placement issues encountered in the development of nested effects, and we briefly described the proposed solution. The aim of this section is to provide practical details on how to implement the constraints on the sample mean and variance. 

The straightforward extension of rescaling the norm of $\bm a$, relevant for linear combinations, to a general transformation $\tilde{s}(\bf{x})$ is
\begin{equation}\label{eq:std_cov_trans}
    \tilde{s}'(\mathbf{x}) = Z[\tilde{s}(\mathbf{x})] = \frac{\tilde{s}({\bf x}) - \bar{\tilde{s}}({\bf x})}{\sqrt{\hat{\var}\left[\tilde{s}({\bf x})\right]}}, 
\end{equation}
where $Z(\cdot)$ denotes the centring and scaling of $\tilde{s}({\bf x})$ by its sample mean and standard deviation. Such a standardisation step can be easily integrated in nested smooth effects. However, note that standardising the transformation can lead to identifiability problems. This is easily seen in the  linear transformation case. In this context, standardisation is equivalent to applying two constraints on $\tilde{s}'(\mathbf{x})$: zero empirical mean and unit norm. Therefore, if $p$ represents the dimension of the vector $\bm a$ of the linear combination coefficients, then the effective number of free parameters is $(p-2)$. Hence, to avoid identifiability issues, two parameters must be removed from $\bm a$. 

In the single index case, standardisation can be achieved via a simple reparametrisation. Let $\bf X$ be the $n\times p$ matrix containing the single index vectors, ${\mathbf{x}}_1, \dots, {\mathbf{x}}_n$, on each row. The first step is to centre the columns of $\bf X$ so that their elements sum to zero. Then, the variance constraint, $\hat{\var}\left(\bm a\ts {\mathbf{x}} \right ) = c$, is satisfied by expressing $\bm a$ as a linear transformation of a unit norm vector $\bm \omega$, i.e. $||\bm \omega|| = 1$. In particular, denote as $\hat{\bm \Sigma}$ the empirical covariance matrix of ${\mathbf{x}}_1, \dots, {\mathbf{x}}_n$. Then ${\bm a} = \sqrt{c}{\bf Q} \bm \Lambda^{-1/2} \bm \omega$, where $\hat{\bm \Sigma} = {\bf Q} \bm \Lambda  {\bf Q}\ts$ and the variance of the linear transformation is then equal to $c$, in fact
\begin{equation*}
       \hat{\var}\left(\bm a\ts {\mathbf{x}} \right ) = \bm a\ts \hat{\bm \Sigma} \bm a = c \bm \omega\ts \bm \Lambda^{-1/2} {\bf Q}\ts {\bf Q} \bm \Lambda  {\bf Q}\ts {\bf Q} \bm \Lambda^{-1/2} \bm \omega = c.
\end{equation*} 




Although this parameterisation bypasses the scaling problem mentioned earlier in this section, it spoils the linearity of the transformation with respect to its parameters. This fact has significant implications: It loses the computational benefits inherent in the linear combination and affects the interpretability of the results due to rotation and scaling via $\sqrt{c}{\bf Q} \bm \Lambda^{-1/2}$. Therefore, to avoid identifiability issues and to preserve the linearity of the effect with respect to the parameters, we adopt a penalty-based approach for all the transformations considered in this work. 


Specifically, define the centred and scaled transformation 
\begin{equation}\label{eq:scaled_nested}
    \tilde{s}'(\mathbf{x}) = e^{a_0}[\tilde{s}(\mathbf{x}) - \bar{\tilde{s}}({\bf x})],
\end{equation}
where the additional parameter $a_0\in \mathbb{R}$ controls the scale of the transformation. Then, the constraint on the variance is imposed by adding a penalty term $q_{\tilde{s}'}= \{\hat{\var}[\tilde{s}'(\mathbf{x})] - c\}^2$ to \eqref{eq:logPosterior}, for each transformation, leading to
\begin{equation} 
    \mathcal{L}(\bm \zeta)
    =\log p(\bm \zeta|{\bm y}, \bm \lambda) 
    = \sum_{i=1}^n \ell_i(\bm \zeta\mid y_i) - \frac{1}{2} \sum_{g=1}^G \lambda_g \bm \zeta \ts {\bf S}_g \bm \zeta- \sum_{u=1}^U q_u(\bm \zeta),
\end{equation}
where $U$ denotes the total number of nested effects in the model. Note that, while the constraint or penalty on $\hat{\text{var}}[\tilde{s}'(\mathbf{x})]$ is imposed on all elements of $\bm a$, in practice only $a_0$ is affected by it. This is because, in the absence of this constraint, $a_0$ would be very weakly identifiable. To see this, consider the case where $s(\tilde{s}'(\mathbf{x}))$ is constructed via an unpenalised polynomial basis, that is, $s[\tilde{s}'(\mathbf{x})] = \tilde{s}'(\mathbf{x})b_1 + [\tilde{s}'(\mathbf{x})]^2b_2 + \cdots + [\tilde{s}'(\mathbf{x})]^Lb_L$. Plugging in $\tilde{s}'(\mathbf{x}) = a_0\tilde{s}(\mathbf{x})$ shows that each power of $a_0$ can be absorbed into the corresponding $b_l$ without changing $s[\tilde{s}'(\mathbf{x})]$. For more complex bases and penalties, the lack of identifiability is less than perfect but it is, in our experience, strong enough to lead to divergence during model fitting, if a constraint is not applied. 

Note that the unidentifiability of $a_0$ is desirable in our context, because it allows us to satisfy the constraint $\hat{\var}[\tilde{s}'(\mathbf{x})] = c$ almost exactly by including the penalty $q_{\tilde{s}'}$ rather than by, for example, reparametrising $\bm a$. Such a reparametrisation would be very cumbersome, because it would need to be analytically derived for each type of transformation (see above for the linear combination case), thus slowing down the development and integration of new transformations. Instead, the penalty-based alternative is agnostic to the type of transformation used, and it is easily integrated into the model fitting framework adopted here. 

The addition of $a_0$ to $\bm a$ and the centring are not necessary for the linear combinations described in Section \ref{sec:linear_proj}. In fact, as explained above, the linear transformation is centred around 0 by subtracting from each column of $\bf X$ its mean, while the scale of $\tilde{s}(\mathbf{x}_i)$ is implicitly controlled by $||\bm a||$. Hence, adding a further scale parameter $a_0$ is unnecessary and $||\bm a||$ is determined by the penalty mentioned above. Finally, note that the sign of $\bm a$ is not identifiable. This is not a problem for model fitting, but some care is needed when interpreting the model output. 

\section{Computational details}
\label{app:derivative_inferno}

\subsection{Log-likelihood derivatives with respect to ${\eta}$
and $\tilde{s}$} \label{app:D_ll_D_eta_s}
To compute the derivative of the log-likelihood w.r.t. the coefficients and smoothing parameters, we apply the chain rule. This requires the partial derivatives of an individual log-likelihood contribution $\ell_i$ w.r.t. the $m$ linear predictors $\bm \eta_i$ and a general nested transformation value $\tilde{s}_i$. Here, we drop the observation index $i$, and we indicate with $\eta_j, \eta_k$, and $\eta_l$ three distinct linear predictors for a single observation, with indices $j, k, l \in \{1, \dots, m\}$ such that $j \neq k$, $k \neq l$ and $j \neq l$. We indicate the corresponding elements of the parameter vector $\bm \theta$ with $\theta_j,\theta_k$, and $\theta_l$. Then, the first-order partial derivative of $\ell$ w.r.t. $\eta_j$ is
\begin{equation*}
    \ell^{\eta_j}=\frac{\partial \ell}{\partial\eta_j}=\ell^{\theta_j}\theta_j^{\eta_j},  
\end{equation*}
where $\theta_j^{\eta_j}=\partial\theta_j/\partial\eta_j$
is the first derivative of the inverse link function. The higher-order derivatives are 
\begin{align*}
    \arraycolsep=1pt\def\arraystretch{1.5}
    \begin{array}{l}
    \ell^{\eta_j\eta_j} = 
    \ell^{\theta_j}\theta_j^{\eta_j\eta_j}+
    \ell^{\theta_j\theta_j}\left(\theta_j^{\eta_j} \right)^{2},\\
    \ell^{\eta_j\eta_k} = 
    \ell^{\theta_j\theta_k}\theta_j^{\eta_j}\theta_k^{\eta_k},
    \end{array} \qquad\qquad
    \begin{array}{l}
        \ell^{\eta_j\eta_j\eta_j} = 
        \ell^{\theta_j}\theta_j^{\eta_j\eta_j\eta_j} + 
        3\ell^{\theta_j\theta_j}\theta_j^{\eta_j}\theta_j^{\eta_j\eta_j} + 
        \ell^{\theta_j\theta_j\theta_j}(\theta_j^{\eta_j})^{3},\\
        \ell^{\eta_j\eta_j\eta_k} = \ell^{\theta_j\theta_j}\theta_j^{\eta_j\eta_j}\theta_k^{\eta_k} + \ell^{\theta_j\theta_j\theta_k} \left(\theta_j^{\eta_j}\right)^2\theta_k^{\eta_k},\\
        \ell^{\eta_j\eta_k\eta_l} = \ell^{\theta_j\theta_k\theta_l} \theta_j^{\eta_j}\theta_k^{\eta_k}\theta_l^{\eta_l}.
    \end{array}
\end{align*}

Consider now three distinct transformations $\tilde{s}_j$, $\tilde{s}_k$ and $\tilde{s}_l$, acting on to linear predictors $\eta_{j^*},\eta_{k^*}$, and $\eta_{l^*}$. The latter could overlap, that is, several transformations could be acting on the same linear predictor (i.e., $j \neq k$ does not imply $j^* \neq k^*$). Then the first-order derivative of the log-likelihood w.r.t $\tilde{s}_j$ is  
\begin{equation*}
   \ell^{\tilde{s}_j}=\frac{\partial \ell}{\partial \tilde{s}_j}=\ell^{\eta_{j^*}}\eta_{j^*}^{\tilde{s}_j}, 
\end{equation*}
where $\eta^{\tilde{s}_j}_{j^*} = \partial\eta_{j^*}/\partial \tilde{s}_j$. Further log-likelihood derivatives w.r.t. $\tilde{s}_j$, $\tilde{s}_k$ and $\tilde{s}_l$ follow the same structure as those w.r.t. $\eta_j, \eta_k$, and $\eta_l$ (provided above), after substituting $\eta$ with $\tilde{s}$, as well as $\theta_{j}$, $\theta_{k}$ and $\theta_{l}$ with $\eta_{j^*}, \eta_{k^*}$, and $\eta_{l^*}$ in each expression. For example, applying the substitutions to $\ell^{\eta_j\eta_k\eta_l} = \ell^{\theta_j\theta_k\theta_l} \theta_j^{\eta_j}\theta_k^{\eta_k}\theta_l^{\eta_l}$ leads to  
$\ell^{\tilde{s}_j\tilde{s}_k\tilde{s}_l} = \ell^{\eta_{j^*}\eta_{k^*}\eta_{l^*}} \eta_{j^*}^{\tilde{s}_j}\eta_{k^*}^{\tilde{s}_k}\eta_{l^*}^{\tilde{s}_l}$. 

The formulas for mixed derivatives w.r.t. both $\eta$ and $\tilde{s}$ can be obtained via those w.r.t. $\tilde{s}$ only. For example, the expression for $\ell^{\eta_{j^*}\tilde{s}_k}$ is the same as that for $\ell^{\tilde{s}_{j}\tilde{s}_k}$, where $\tilde{s}_{j}$ is an identity transformation acting on the $j^*$-th linear predictor. That is, $\tilde{s}_{j} = \eta_{j^*}$, so that $\eta^{\tilde{s}_j}_{j^*} = 1$ and  $\eta^{\tilde{s}_j\tilde{s}_j}_{j^*} =  \eta^{\tilde{s}_j\tilde{s}_j\tilde{s}_j}_{j^*} = 0$. 
Instead, for general transformations, $\eta^{\tilde{s}_j}_{j^*}=({\bf m}^1_{\bf b})\ts{\bf b}$, with ${\bf m}^1_{\bf b}$ representing the relevant row of ${\bf M}^1_{\bf b}$ and $\bf b$ being the spline coefficients of the corresponding nested effect. Second, $\eta^{\tilde{s}_j\tilde{s}_j}_{j^*}$, and third  $\eta^{\tilde{s}_j\tilde{s}_j\tilde{s}_j}_{j^*}$ derivatives simply require substituting ${\bf m}^1_{\bf b}$ with ${\bf m}^2_{\bf b}$ and ${\bf m}^3_{\bf b}$, respectively. For effects based on linear combinations ${\bf b}$ should be substituted with $\bm \beta$.

Finally, note that the system of log-likelihood derivatives w.r.t. the linear predictors is simplified by the fact that $\theta_j^{\eta_j\eta_k} = \theta_j^{\eta_j\eta_k\eta_l} = 0$ unless $j = k = l$, due to the one-to-one relation between linear predictors and distributional parameters. Similarly, in the derivatives w.r.t the transformations, we have that $\eta_{j^*}^{\tilde{s}_j\tilde{s}_k} = \eta_{j^*}^{\tilde{s}_j\tilde{s}_k\tilde{s}_l} = 0$ unless $j = k = l$, even when the $\tilde{s}_j$, $\tilde{s}_k$ and $\tilde{s}_l$ act on the same linear predictor (i.e., $j^* = k^* = l^*$).

\subsection{Exceptions to the derivatives of Hessian blocks}\label{app:exceptions_dH}
Here we provide the derivatives of the Hessian blocks w.r.t. $\rho$ that do not follow the standard pattern $\bm{\Gamma}_{\bm{\psi}_3}^{\bm{\psi}_1\bm{\psi}_2}={\bf M}_{\bm{\psi}_{2}}\ts{\bf V}_{\bm{\psi}_{3}}^{\bm{\psi}_{1}\bm{\psi}_{2}}{\bf M}_{\bm{\psi}_{1}}$. To do so, we need to expand the notation defined in Section~\ref{sec:deriv_framework}. In particular, all the exceptional $\bm{\Gamma}_{\bm{\psi}_3}^{\bm{\psi}_1\bm{\psi}_2}$ terms involve triplets $\{\bm{\psi}_1,\bm{\psi}_2, \bm{\psi}_3\}$ where at least two of the three vectors control the same nested smooth effect. In fact, $\bm{\Gamma}_{\bm{\psi}_3}^{\bm{\psi}_1\bm{\psi}_2}$ terms where each of three vectors in $\{\bm{\psi}_1,\bm{\psi}_2, \bm{\psi}_3\}$ controls a different effects follow the standard pattern \eqref{eq:general_block}. Among the exceptional blocks, additional notation is needed for cases where two of the vectors in $\{\bm{\psi}_1,\bm{\psi}_2, \bm{\psi}_3\}$ control the same nested effect, while the third controls a different effect. In particular, we expand the definition of a generic parameter vector from $\bm \psi \in \{\bm \gamma,{\bm a}, {\bm b},  \bm \alpha, \bm \beta\}$ (as in Section~\ref{sec:deriv_framework}), to $\bm \psi \in \{\bm \gamma,{\bm a}, {\bm b},  \bm \alpha, \bm \beta,{\bm a}^*, {\bm b}^*,  \bm \alpha^*, \bm \beta^*\}$, where the superscript $*$ is used to indicate that $\{{\bm a}, {\bm b}\}$ and  $\{{\bm a}^*, {\bm b}^*\}$ control different nested effects (same for $\{\bm \alpha, \bm \beta\}$ and $\{\bm \alpha^*,\bm \beta^*\}$).

Assume for simplicity that $\text{dim}(\bm \gamma) = \text{dim}(\bm \alpha) = \dots = p$. As detailed below, terms $\bm{\Gamma}_{\bm{\psi}_3}^{\bm{\psi}_1\bm{\psi}_2}$ where more than one vector in $\{\bm{\psi}_1,\bm{\psi}_2, \bm{\psi}_3\}$ is of class $\bm a$ will require computing $\partial^2\tilde{s}_i/\partial a_j\partial a_k$, while terms of type $\bm{\Gamma}_{\bm{a}}^{\bm{a}\bm{a}}$ involve computing $\partial^3\tilde{s}_i/\partial a_j\partial a_k\partial a_l$, for $k$, $j$ and $l = 1, \dots, p$, with $i = 1, \dots, n$. Hence, terms of type $\bm{\Gamma}_{\bm{a}}^{\bm{a}\bm{a}}$ are the only ones requiring $O(np^3)$ computation, all other exceptional terms having cost $O(np^2)$, as for the terms following the standard pattern. 

The expressions for the terms $\bm{\Gamma}_{\bm{\psi}_3}^{\bm{\psi}_1\bm{\psi}_2}$ involving two or three vectors of class $\bm a$ require further additions to the notation provided in Section~\ref{sec:deriv_framework}. Recall that ${\bf M}_{\bm a}$ and ${\bf M}_{\bm \alpha}$ are matrices such that ${\bf M}_{\bm a} = \nabla_{\bm{a}}\ts{\tilde{\bf s}}$ and ${\bf M}_{\bm \alpha} = \nabla_{\bm \alpha}\ts{\tilde{\bf s}}$, where $\nabla_{\bm{a}}\ts\tilde{\bf s} = \ilpdif{\tilde{\bf s}}{\bm{a}\ts}$  and $\tilde{\bf s}$ is the vector containing the $n$ observed values of a transformation $\tilde{s}$. For linear combinations, further derivatives of $\tilde{\bf s}$ w.r.t. $\bm \alpha$ are equal to zero, while for general transformations they are stored in tensors with elements
\begin{align*}
({\bf M}^1_{\bm a})_{ijk} = \frac{\partial^2 \tilde{s}_i}{\partial a_j\partial a_k}, &&
({\bf M}^2_{\bm a})_{ijkl} = \frac{\partial^3 \tilde{s}_i}{\partial a_j\partial a_k\partial a_l}.
\end{align*}
Such derivatives are used to compute the tensors
\begin{align*}
({\bf P}_{\bm{a}}^1)_{ij}=\sum_{k=1}^p ({\bf M}^1_{\bm a})_{ijk}\dif{\hat{a}_k}{\rho}, &&
({\bf P}_{\bm{a}}^2)_{ijk}=\sum_{l=1}^p ({\bf M}^2_{\bm a})_{ijkl}\dif{\hat{a}_l}{\rho}.
\end{align*}
which appear in several places below.

Having extended the notation from Section~\ref{sec:deriv_framework}, the exceptional terms are:
\begin{itemize}
	\item $\bm{\Gamma}_{\bm{\beta}}^{\bm{\alpha}\bm{\alpha}}$, where the diagonal matrix has two additional terms 
	\begin{equation*}
		\left({\bf V}_{\bm{\beta}}^{\bm{\alpha}\bm{\alpha}}\right)_{ii} 
		=  \ell_{i}^{\nu(\bm{\alpha}, \bm{\alpha}, \bm{\beta})}\nu_{\bm{\beta}}(\bm{\beta})_{i} 
		+2\ell_{i}^{\nu(\bm{\alpha}, \bm{\beta})}\nu_{\bm{\beta}}^{1}(\bm{\beta})_{i}+\ell_{i}^{\nu(\bm{\beta})}\nu_{\bm{\beta}}^{2}(\bm{\beta})_{i}.
	\end{equation*}
	\item $\bm{\Gamma}_{\bm{\alpha}}^{\bm{\beta}\bm{\beta}}$ (and $\bm{\Gamma}_{\bm{a}}^{\bm{b}\bm{b}}$) have two additional terms
	\begin{equation*}
		\bm{\Gamma}_{\bm{\alpha}}^{\bm{\beta}\bm{\beta}} 
		= \left({{\bf M}_{\bm{\beta}}}\right)\ts{\bf V}_{\bm{\alpha}1}^{\bm{\beta}\bm{\beta}}{\bf M}_{\bm{\beta}} 
		+ \left({{\bf M}_{\bm{\beta}}^{1}}\right)\ts{\bf V}_{\bm{\alpha}2}^{\bm{\beta}\bm{\beta}}{\bf M}_{\bm{\beta}} 
		+ \left[\left({{\bf M}_{\bm{\beta}}^1}\right)\ts{\bf V}_{\bm{\alpha}2}^{\bm{\beta}\bm{\beta}}{\bf M}_{\bm{\beta}}\right]\ts,	
	\end{equation*}
	where 
	\begin{align*}
		\left({\bf V}_{\bm{\alpha}1}^{\bm{\beta}\bm{\beta}}\right)_{ii}
		= \ell_{i}^{\nu(\bm{\beta}, \bm{\beta}, \bm{\alpha})}\nu_{\bm{\alpha}}(\bm{\alpha})_{i},
		&&
		\left({\bf V}_{\bm{\alpha}2}^{\bm{\beta}\bm{\beta}}\right)_{ii}
		= \ell_{i}^{\nu(\bm{\beta}, \bm{\beta})}\nu_{\bm{\alpha}}(\bm{\alpha})_{i}.
	\end{align*}
	The expression for $\bm{\Gamma}_{\bm{a}}^{\bm{b}\bm{b}}$ is obtained substituting $\bm{\beta}$ and $\bm{\alpha}$ with $\bm{b}$ and $\bm{a}$, respectively.
	\item $\bm{\Gamma}_{\bm{\psi}_{3}}^{\bm{\alpha}\bm{\beta}}$ (or $\big(\bm{\Gamma}_{\bm{\psi}_{3}}^{\bm{\beta}\bm{\alpha}}\big)\ts$)
	has seven cases. If $\bm{\psi}_{3}\in \left\{\bm{\gamma}, \bm{\alpha}_{*}, \bm{\beta}_{*}, \bm{a}, \bm{b}\right\}$, we have
	\begin{equation*}
		\bm{\Gamma}_{\bm{\psi}_{3}}^{\bm{\alpha}\bm{\beta}}
		= {\bf M}_{\bm{\beta}}\ts{\bf V}_{\bm{\psi}_{3}}^{\bm{\alpha}\bm{\beta}}{\bf M}_{\bm{\alpha}}
		+\left({{\bf M}_{\bm{\beta}}^1}\right)\ts{\bf D}_{\bm{\psi}_{3}}{\bf M}_{\bm{\alpha}},
	\end{equation*}
	where 
	\begin{equation*}
		({\bf D}_{\bm{\psi}_{3}})_{ii} 
		= \ell_{i}^{\nu(\bm{\beta}, \bm{\psi}_{3})}\nu_{\bm{\psi}_{3}}(\bm{\psi}_{3})_i.	
	\end{equation*}
	If $\bm{\psi}_{3}\in \{\bm{\alpha}, \bm{\beta}\}$ we have
	\begin{equation*}
		\bm{\Gamma}_{\bm{\alpha}}^{\bm{\alpha}\bm{\beta}} + \bm{\Gamma}_{\bm{\beta}}^{\bm{\alpha}\bm{\beta}} 
		= \left[{{\bf M}_{\bm{\beta}}}\ts{\bf V}_{\bm{\alpha}\bm{\beta}1}^{\bm{\alpha}\bm{\beta}}
		+ ({\bf M}_{\bm{\beta}}^{1})\ts{\bf V}_{\bm{\alpha}\bm{\beta}2}^{\bm{\alpha}\bm{\beta}}
		+ ({\bf M}_{\bm{\beta}}^{2})\ts{\bf V}_{\bm{\alpha}\bm{\beta}3}^{\bm{\alpha}\bm{\beta}}
		\right]{\bf M}_{\bm{\alpha}},
	\end{equation*}
	where 
	\begin{equation*}
		\left({\bf V}_{\bm{\alpha}\bm{\beta}1}^{\bm{\alpha}\bm{\beta}}\right)_{ii}
		= \ell_{i}^{\nu(\bm{\beta}, \bm{\beta}, \bm{\alpha})}\nu_{\bm \beta}(\bm{\beta})
		+\ell_{i}^{\nu(\bm{\beta}, \bm{\beta})}p^1_{\bm \beta}(\bm{\beta})
		+\ell_{i}^{\nu(\bm{\beta}, \bm{\alpha}, \bm{\alpha})} \nu_{\bm{\alpha}}(\bm{\alpha}),
	\end{equation*}
	\begin{align*}
		\left({\bf V}_{\bm{\alpha}\bm{\beta}2}^{\bm{\alpha}\bm{\beta}}\right)_{ii}
		= \ell_{i}^{\nu(\bm{\beta}, \bm{\beta})}\nu_{\bm \beta}(\bm{\beta})
		 +2\ell_{i}^{\nu(\bm{\beta}, \bm{\alpha})} \nu_{\bm{\alpha}}(\bm{\alpha}),
		&&
		({\bf V}_{\bm{\alpha}\bm{\beta}3}^{\bm{\alpha}\bm{\beta}})_{ii}
		 = \ell_{i}^{\nu(\bm{\beta})}\nu_{\bm{\alpha}}(\bm{\alpha}).
	\end{align*}
	\item $\bm{\Gamma}_{\bm{\psi}_{3}}^{\bm{a}\bm{b}}$ (or $(\bm{\Gamma}_{\bm{\psi}_{3}}^{\bm{b}\bm{a}})\ts$)
	has seven cases. When $\bm{\psi}_{3}\in \{\bm{\gamma}, \bm{\alpha}, \bm{\beta}, \bm{a}_{*}, \bm{b}_{*}\}$ it is analogous to $\bm{\Gamma}_{\bm{\psi}_{3}}^{\bm{\alpha}\bm{\beta}}$
	(with $\bm{\psi}_{3}\in \{\bm{\gamma}, \bm{\alpha}, \bm{\beta}, \bm{a}_{*}, \bm{b}_{*}\}$).
	The cases $\bm{\psi}_{3}=\bm{a}$ or $\bm{b}$ are harder, that is,
	\begin{equation*}
		\bm{\Gamma}_{\bm{b}}^{\bm{a}\bm{b}} 
		= \left[{\bf M}_{\bm{b}}\ts{\bf V}_{\bm{b}1}^{\bm{a}\bm{b}}
		+({\bf M}_{\bm{b}}^{1})\ts{\bf V}_{\bm{b}2}^{\bm{a}\bm{b}}
		\right]{\bf M}_{\bm{a}},
	\end{equation*}
	where 
	\begin{align*}
		\left({\bf V}_{\bm{b}1}^{\bm{a}\bm{b}}\right)_{ii} 
		= \ell_{i}^{\nu(\bm{a}, \bm{b}, \bm{b})}\nu_{\bm b}(\bm{b})
		+\ell_{i}^{\nu(\bm{b}, \bm{b})}p^1_{\bm b}(\bm{b})
		&&
		\left({\bf V}_{\bm{b}2}^{\bm{a}\bm{b}}\right)_{ii}
		= \ell_{i}^{\nu(\bm{b}, \bm{b})}\nu_{\bm b}(\bm{b}).
	\end{align*}
	While
	\begin{equation*}
		\bm{\Gamma}_{\bm{a}}^{\bm{a}\bm{b}}
		= \left[{\bf M}_{\bm{b}}\ts{\bf V}_{\bm{a}1}^{\bm{a}\bm{b}}
		  + ({\bf M}_{\bm{b}}^{1})\ts{\bf V}_{\bm{a}2}^{\bm{a}\bm{b}}
		  + ({\bf M}_{\bm{b}}^{2})\ts{\bf V}_{\bm{a}3}^{\bm{a}\bm{b}}\right]{\bf M}_{\bm{a}}
		  + \left[{\bf M}_{\bm{b}}\ts
		  + ({\bf M}_{\bm{b}}^{1})\ts\right]{\bf V}_{\bm{a}4}^{\bm{a}\bm{b}}{\bf P}^1_{\bm a},
	\end{equation*}
	where
	\begin{align*}
		\left({\bf V}_{\bm{a}1}^{\bm{a}\bm{b}}\right)_{ii} 
		&= \ell_{i}^{\nu(\bm{b}, \bm{a}, \bm{a})}\nu_{\bm{a}}(\bm{a})_i,
		&
		\left({\bf V}_{\bm{a}2}^{\bm{a}\bm{b}}\right)_{ii}
		&= 2\ell_{i}^{\nu(\bm{b}, \bm{a})}\nu_{\bm{a}}(\bm{a})_i\\
		\left({\bf V}_{\bm{a}3}^{\bm{a}\bm{b}}\right)_{ii} 
		&= \ell_{i}^{\nu(\bm{b})}\nu_{\bm{a}}(\bm{a})_i,
		&
		\left({\bf V}_{\bm{a}4}^{\bm{a}\bm{b}}\right)_{ii}
		&= \ell_{i}^{\nu(\bm{b}, \bm{a})}+\ell_{i}^{\nu(\bm{b})}
	\end{align*}
	\item $\bm{\Gamma}_{\bm{\psi}_{3}}^{\bm{a}\bm{a}}$ has seven cases. When $\bm{\psi}_{3}=\bm{a}$ it must be computed via
	\begin{equation*}
		\bm{\Gamma}_{\bm{a}}^{\bm{a}\bm{a}} 
		= {\bf M}_{\bf a}\ts{\bf V}_{\bm{a}1}^{\bm{a}\bm{a}}{\bf M}_{\bf a}
		+{\bf M}_{\bf a}\ts{\bf V}_{\bm{a}2}^{\bm{a}\bm{a}}{\bf P}^1_{\bm a}
		+\left({\bf M}_{\bf a}\ts{\bf V}_{\bm{a}2}^{\bm{a}\bm{a}}{\bf P}^1_{\bm a}\right)\ts
		+{\bf V}_{\bm{a}3}^{\bm{a}\bm{a}}
		+{\bf V}_{\bm{a}4}^{\bm{a}\bm{a}}		
	\end{equation*}
	where
	\begin{align*}
		\left({\bf V}_{\bm{a}1}^{\bm{a}\bm{a}}\right)_{ii}
		&= \ell_{i}^{\nu(\bm{a},\bm{a},\bm{a})}\nu_{\bm a}(\bm{a})_i,
		&	
		\left({\bf V}_{\bm{a}2}^{\bm{a}\bm{a}}\right)_{ii}
		&= \ell_{i}^{\nu(\bm{a},\bm{a})},\\
		\left({\bf V}_{\bm{a}3}^{\bm{a}\bm{a}}\right)_{jk} 
		&= \sum_{i=1}^n\left[\ell_{i}^{\nu(\bm{a},\bm{a})}\nu_{\bm a}(\bm{a})_i\right]\left({\bf M}_{\bm{a}}^1\right)_{ijk},
		&
		\left({\bf V}_{\bm{a}4}^{\bm{a}\bm{a}}\right)_{jk}
		&= \sum_{i=1}^n\ell_{i}^{\nu(\bm{a})}\left({\bf P}_{\bm{a}}^2\right)_{ijk}.
	\end{align*}
    The case where $\bm{\psi}_{3}=\bm{b}$ is
	\begin{equation*}
		\bm{\Gamma}_{\bm{b}}^{\bm{a}\bm{a}}={\bf M}_{\bf a}\ts{\bf V}_{\bm{b}1}^{\bm{a}\bm{a}}{\bf M}_{\bf a}+{\bf V}_{\bm{b}2}^{\bm{a}\bm{a}},		
	\end{equation*}
	where
	\begin{align*}
		\left({\bf V}_{\bm{b}1}^{\bm{a}\bm{a}}\right)_{ii}
		& = \ell_{i}^{\nu(\bm{a}, \bm{a}, \bm{b})}\nu_{\bm{b}}(\bm{b})_{i} 
			+2\ell_{i}^{\nu(\bm{a}, \bm{b})}\nu_{\bm{b}}^{1}(\bm{b})_{i}
			+\ell_{i}^{\nu(\bm{b})}\nu_{\bm{b}}^{2}(\bm{b})_{i},\\
		    \left({\bf V}_{\bm{b}2}^{\bm{a}\bm{a}}\right)_{jk}
		& = \sum_{i=1}^n\left[\ell_{i}^{\nu(\bm{a}, \bm{b})}\nu_{\bm b}(\bm{b})_i
		    + \ell_{i}^{\nu(\bm{b})}\nu_{\bm{b}}^{1}(\bm{b})_{i}\right]\left({\bf M}_{\bm{a}}^1\right)_{ijk}.
	\end{align*}
	The case where $\bm{\psi}_{3} \notin \{\bm{a},\bm{b}$\} is
	\begin{equation*}
		\bm{\Gamma}_{\bm{\psi_{3}}}^{\bm{a}\bm{a}} 
		= {\bf M}_{\bf a}\ts{\bf V}_{\bm{\psi}_{3}1}^{\bm{a}\bm{a}}{\bf M}_{\bf a}+{\bf V}_{\bm{\psi}_{3}2}^{\bm{a}\bm{a}},	
	\end{equation*}
	where
	\begin{align*}
		\left({\bf V}_{\bm{\psi}_{3}1}^{\bm{a}\bm{a}}\right)_{ii} 
		= \ell_{i}^{\nu(\bm{a}, \bm{a}, \bm{\psi}_{3})},
		&&
		\left({\bf V}_{\bm{\psi}_{3}2}^{\bm{a}\bm{a}}\right)_{jk}
		= \sum_{i=1}^n\left[\ell_{i}^{\nu(\bm{a}, \bm{\psi}_{3})}\nu_{\bm{\psi}_3}(\bm{\psi}_3)_i\right]\left({\bf M}_{\bm{a}}^1\right)_{ijk}.
	\end{align*}
	\item $\bm{\Gamma}_{\bm{a}}^{\bm{a}\bm{\psi}_{2}}$ with $\bm{\psi}_{2} \notin \{\bm{a},\bm{b}\}$
	is 
	\begin{equation*}
		\bm{\Gamma}_{\bm{a}}^{\bm{a}\bm{\psi}_{2}}={\bf M}_{\bm{\psi}_{2}}\ts{\bf V}_{\bm{a}1}^{\bm{a}\bm{\psi}_{2}}{\bf M}_{\bm{\psi}_{1}}+{\bf M}_{\bm{\psi}_{2}}\ts{\bf V}_{\bm{a}2}^{\bm{a}\bm{\psi}_{2}}{\bf P}_{\bm{a}}^1,		
	\end{equation*}
	where
	\begin{align*}
		\left({\bf V}_{\bm{a}1}^{\bm{a}\bm{\psi}_{2}}\right)_{ii}
		= \ell_{i}^{\nu(\bm{a}, \bm{a}, \bm{\psi}_{2})}\nu_{\bm{a}}(\bm{a})_{i},
		&&
		({\bf V}_{\bm{a}2}^{\bm{a}\bm{\psi}_{2}})_{ii}=\ell_{i}^{\nu(\bm{a},\bm{\psi}_{2})}.
	\end{align*}
	\item For the triplets $\bm{\Gamma}_{\bm{\beta}}^{\bm{\psi}_1\bm{\alpha}}$ (or $(\bm{\Gamma}_{\bm{\beta}}^{\bm{\alpha}\bm{\psi}_1})\ts$) with $\bm{\psi}_1 \notin \{\bm{\alpha},\bm{\beta}\}$ and $\bm{\Gamma}_{\bm{b}}^{\bm{\psi}_1\bm{a}}$ (or $(\bm{\Gamma}_{\bm{b}}^{\bm{a}\bm{\psi}_1})\ts$) with $\bm{\psi}_1 \notin \{\bm{a},\bm{b}\}$ we have 
	\begin{equation*}
		\bm{\Gamma}_{\bm{\beta}}^{\bm{\psi}_{1}\bm{\alpha}}
		= {\bf M}_{\bm{\alpha}}\ts\left({\bf V}_{\bm{\beta}}^{\bm{\psi}_{1}\bm{\alpha}}
		+{\bf E}_{\bm{\beta}}^{\bm{\psi}_{1}}\right){\bf M}_{\bm{\psi}_{1}},		
	\end{equation*}
	with
	\begin{align*}
    \left({\bf V}_{\bm{\beta}}^{\bm{\psi}_{1}\bm{\alpha}}\right)_{ii}
		= \ell_{i}^{\nu(\bm{\psi_1}, \bm{\alpha}, \bm{\beta})}\nu_{\bm{\beta}}(\bm{\beta})_{i}, &&
		\left({\bf E}_{\bm{\beta}}^{\bm{\psi}_{1}}\right)_{ii}
		=\ell_{i}^{\nu(\bm{\psi}_{1}, \bm{\beta})}\nu_{\bm{\beta}}^{1}(\bm{\beta})_i.		
	\end{align*}
    
	The triplets of the form $\bm{\Gamma}_{\bm{\alpha}}^{\bm{\psi}_1\bm{\beta}}$ (or $(\bm{\Gamma}_{\bm{\alpha}}^{\bm{\beta}\bm{\psi}_1})\ts$), with $\bm{\psi}_1 \notin \{\bm{\alpha},\bm{\beta}\}$, 	and $\bm{\Gamma}_{\bm{a}}^{\bm{\psi}_1\bm{b}}$ (or $(\bm{\Gamma}_{\bm{a}}^{\bm{b}\bm{\psi}_1})\ts$), with $\bm{\psi}_1 \notin \{\bm{a},\bm{b}\}$, can be computed as follows 
	\begin{equation*}
		\bm{\Gamma}_{\bm{\alpha}}^{\bm{\psi}_{1}\bm{\beta}}
		= {\bf M}_{\bm{\beta}}\ts{\bf V}_{\bm{\alpha}}^{\bm{\psi}_{1}\bm{\beta}}{\bf M}_{\bm{\psi}_{1}}
		+{\bf G}_{\bm{\alpha}}^{\bm{\psi}_{1}\bm{\beta}},		
	\end{equation*}
	where
	\begin{align*}
    \left({\bf V}_{\bm{\alpha}}^{\bm{\psi}_{1}\bm{\beta}}\right)_{ii} = \ell_{i}^{\nu(\bm{\psi_1}, \bm{\beta}, \bm{\alpha})}\nu_{\bm{\alpha}}(\bm{\alpha})_{i}, &&
		{\bf G}_{\bm{\alpha}}^{\bm{\psi}_{1}\bm{\beta}}
		= \left({\bf M}_{\bm{\beta}}^{1}\right)\ts{\bf F}_{\bm{\alpha}}^{\bm{\psi}_{1}\bm{\beta}}{\bf M}_{\bm{\psi}_{1}},		
		&&
		\left({\bf F}_{\bm{\alpha}}^{\bm{\psi}_{1}\bm{\beta}}\right)_{ii}
		= \ell_{i}^{\nu(\bm{\psi}_{1}, \bm{\beta})}\nu_{\bm{\alpha}}(\bm{\alpha})_i.
	\end{align*}
    The formula for $\bm{\Gamma}_{\bm{b}}^{\bm{\psi}_1\bm{a}}$ and $\bm{\Gamma}_{\bm{a}}^{\bm{\psi}_1\bm{b}}$ are obtained by substituting $\bm{\alpha}$ and $\bm{\beta}$ with $\bm{a}$ and $\bm{b}$, respectively.
\end{itemize}

\subsection{Derivatives of a transformation $\tilde{s}$ w.r.t. its parameters {\boldmath $a$}}

If $\tilde{s}$ is a linear combination, then all derivatives of order higher than 1 are zero, and the gradient is simply the matrix of inner covariates, i.e., $\pdif{\tilde{s}(\bf x)}{\bm a} = \bf x$. For the remainder of this section, we will focus on the case where $\tilde{s}$ is an adaptive exponential smoothing or a multivariate kernel smoothing.

In Section~\ref{app:scaling_penalty}, a centred and scaled nested transformation is defined by
\begin{equation}
    \tilde{s}'(\mathbf{x}) = e^{a_0}[\tilde{s}(\mathbf{x}) - \bar{\tilde{s}}({\bf x})].
\end{equation}
By differentiating this equation, we obtain:
\begin{align*}
    \pdif{\tilde{s}'({\bf x})}{a_j}=&
    \begin{cases}
        \tilde{s}'({\bf x}) & j=0\\
        e^{a_0}\left[\pdif{\tilde{s}({\bf x})}{a_j} - n^{-1}\sum_{i=1}^n \pdif{\tilde{s}({\bf x}_i)}{a_j}\right] & \text{otherwise}
    \end{cases}\\
    \pddif{\tilde{s}'(\bf{x})}{a_j}{a_k}=&
    \begin{cases}
        \pdif{\tilde{s}'(\bf{x})}{a_j} & k=0\\
        e^{a_0}\left[\pddif{\tilde{s}({\bf x})}{a_j}{a_k} - n^{-1}\sum_{i=1}^n \pddif{\tilde{s}({\bf x}_i)}{a_j}{a_k}\right] & \text{otherwise}
    \end{cases}\\
    \ptdif{\tilde{s}'(\bf{x})}{a_j}{a_k}{a_l}=&
    \begin{cases}
        \pddif{\tilde{s}'(\bf{x})}{a_j}{a_k} & l=0\\
        e^{a_0}\left(\ptdif{\tilde{s}(\bf{x})}{a_j}{a_k}{a_l} - n^{-1}\sum_{i=1}^n \ptdif{\tilde{s}(\bf{x}_i)}{a_j}{a_k}{a_l}\right) & \text{otherwise}
    \end{cases}.
\end{align*}
It is important to note that all the expressions above are written in terms of the derivatives of $\tilde{s}$. In the following paragraphs, we provide the formulas for computing $\partial{\tilde{s}}/\partial {a_j}$, $\partial^2{\tilde{s}}/\partial a_j \partial a_k$, and $\partial^3{\tilde{s}}/\partial a_j\partial a_k\partial a_l$, for the two types of (non-linear) nested transformations considered here.

\subsubsection{Adaptive exponential smoothing}

Recall the adaptive exponential smoothing definition given in Section~\ref{sec:exp_smooth}:
\begin{equation*}
    \tilde{s}(x_i) = \tilde{s}_i = 
    \begin{cases}
       \omega_{i}\tilde{s}_{i-1}+(1-\omega_{i})x_i & i>1\\
       \omega_{i}x_0+(1-\omega_{i})x_{1} & i=1
    \end{cases},
\end{equation*}
where $x_{0}$ is fixed to some value and $\omega_{i}\in(0,1)$. The $\omega_{i}'s$
are modelled by $\omega_{i}=\phi(\tilde{{\bf x}}_{i}\ts\bm{a})$ where $\phi(\cdot)$ is the logistic function. To simplify the notation, denote the derivative with respect to the $j$-th parameter in vector $\bm{a}$ using a superscript, e.g., $\tilde{\bm s}^j= \partial\tilde{\bm s}/{\partial a_j}$. Assuming that $x_{0}$ is fixed and known, the derivatives of $\tilde{s}_{i}$ w.r.t.
$a_{j}$ are
\begin{equation*}
    \tilde{s}_{i}^{j}=
    \begin{cases}
        \omega_{i}^{j}(\tilde{s}_{i-1}-x_i)+\omega_{i}\tilde{s}_{i-1}^{j} & i>1\\
        \omega_{1}^{j}(x_{0}-x_{1}) & i=1
    \end{cases},
\end{equation*}
\begin{equation*}
    \tilde{s}_{i}^{jk}=
    \begin{cases}
        \omega_{i}^{jk}(\tilde{s}_{i-1}-x_i)+\omega_{i}^{j}\tilde{s}_{i-1}^{k}+\omega_{i}^{k}\tilde{s}_{i-1}^{j}+\omega_{i}\tilde{s}_{i-1}^{jk} & i>1\\
        \omega_{1}^{jk}(x_{0}-x_{1}) & i=1
    \end{cases},
\end{equation*}
\begin{equation*}
    \tilde{s}_{i}^{jkl} =
    \begin{cases}
    \begin{aligned}
    &\omega_{i}^{jkl}(\tilde{s}_{i-1}-x_i) + \omega_{i}^{jk}\tilde{s}_{i-1}^{l}
     + \omega_{i}^{jl}\tilde{s}_{i-1}^{k} + \omega_{i}^{kl}\tilde{s}_{i-1}^{j}+ \\
    &\hspace{4cm} + \omega_{i}^{j}\tilde{s}_{i-1}^{kl} + \omega_{i}^{k}\tilde{s}_{i-1}^{jl}
     + \omega_{i}^{l}\tilde{s}_{i-1}^{jk} + \omega_{i}\tilde{s}_{i-1}^{jkl} 
    \end{aligned}
    & i>1\\[6pt]
    \omega_{1}^{jkl}(x_{0}-x_{1}) & i=1
    \end{cases}.
\end{equation*}
where
\begin{align*}
    \omega_{i}^{j}=\phi'_{i}\tilde{x}_{ij},
    &&
    \omega_{i}^{jk} = \phi''_{i}\tilde{x}_{ij}\tilde{x}_{ik},
    &&
    \omega_{i}^{jkl} = \phi'''_{i}\tilde{x}_{ij}\tilde{x}_{ik}\tilde{x}_{il}
\end{align*}
and $\phi', \phi''$ and $\phi'''$ correspond to the first-, second-, and third-order derivatives of the logistic function, respectively.

\subsubsection{Multivariate kernel smoothing}
Let $K_{\bm a}$ be a multivariate kernel density function parameterized by the vector ${\bm a}$, and $z_{ij}$ be a scalar covariate corresponding to the $d$-dimensional vector ${{\bf x}_{ij}}$. A multivariate kernel smoothing transformation, defined in \eqref{eq:mgks}, can be written
\begin{align*}
    \tilde{s}_{i}=
    \bm{\kappa}_{i}\ts {\bf z}_{i}=
    \frac{\sum_{u\in \mathcal{N}_i} K_{\bm a}({\bf x}_{i},{\bf x}_{iu})z_{iu}}{\sum_{q\in \mathcal{N}_i} K_{\bm{a}}({\bf x}_{i},{\bf x}_{iq})}
    && \text{with} &&
    \kappa_{iu} = \frac{K_{\bm a}({\bf x}_{i},{\bf x}_{iu})}{\sum_{q\in \mathcal{N}_i} K_{\bm{a}}({\bf x}_{i},{\bf x}_{iq})}.
\end{align*}
The point at which we evaluate the smooth is ${\bf x}_{i}$. We simplify notation by removing index $i$ and the dependency of $K$ on $\bm{a}$, so that $K_{u} = K_{\bm a}({\bf x}_{i},{\bf x}_{iu})$, and by denoting $\sum_{u\in \mathcal{N}_i}$ with $\sum_{u}$. The 
derivative of $\kappa_u$ w.r.t. $a_j$ is 
\begin{align*}
    \kappa_{u}^{j}
    &=\pdif{\kappa_u}{a_j}
    = \frac{K_{u}^{j}\sum_{f}K_{f}-K_{u}\sum_{f}K_{f}^{j}}{(\sum_{q}K_{q})^{2}}
    =\frac{K_{u}^{j}}{\sum_{q}K_{q}}-\kappa_{u} \frac{\sum_{f}K_{f}^{j}}{\sum_{q}K_{q}}\\
    &=\frac{K_{u}^{j}}{K_{u}}\kappa_{u}-\kappa_{u}\sum_{f}\left(\frac{K_{f}^{j}}{\sum_{q}K_{q}}\frac{K_{f}}{K_{f}}\right)
    =L_{u}^{j}\kappa_{u}-\kappa_{u}\sum_{f}\left(\frac{\kappa_{f}K_{f}^{j}}{K_{f}}\right)\\
    &=\kappa_{u}\left(L_{u}^{j}-\sum_{f}\kappa_{f}L_{f}^{j}\right),
\end{align*}
where $K_{u}^{j} = \partial K_{u}/ \partial a_j$ and $L_{u}=\log K_{u}$, implying that $L_{u}^j = \partial L_{u}/ \partial a_j = K_{u}^{j}/K_{u}$.
The second- and third-order derivatives w.r.t. the elements of $\bm a$ are
\begin{equation*}
    \kappa_{u}^{jk}=
    \kappa_{u}^{k}\left(L_{u}^{j}-\sum_{f}\kappa_{f}L_{f}^{j}\right) +\kappa_{u}\left(L_{u}^{jk}-\sum_{f}\left[\kappa_{f}^{k}L_{f}^{j}+\kappa_{f}L_{f}^{jk}\right)\right],
\end{equation*}
and
\begin{align*}
    \kappa_{u}^{jkl}
    &= \kappa_{u}^{kl}\left(L_{u}^{j}-\sum_{f}\kappa_{f}L_{f}^{j}\right)
    +\kappa_{u}^{k}\left[L_{u}^{jl}-\sum_{f}\left(\kappa_{f}^{l}L_{f}^{j}+\kappa_{f}L_{f}^{jl}\right)\right] +\\
    &\qquad +\kappa_{u}^{l}\left[L_{u}^{jk}-\sum_{f}\left(\kappa_{f}^{k}L_{f}^{j}+\kappa_{f}L_{f}^{jk}\right)\right]+\\
    &\qquad +\kappa_{u}\left[L_{u}^{jkl}-\sum_{f}\left(\kappa_{f}^{kl}L_{f}^{j}+\kappa_{f}^{k}L_{f}^{jl}+\kappa_{f}^{l}L_{f}^{jk}+\kappa_{f}L_{f}^{jkl}\right)\right].
\end{align*}

The general formulas above apply to any  sufficiently differentiable kernel. They can be simplified if $K_{\bm a}$ is an unnormalized Gaussian kernel, with diagonal covariance matrix, that is
$$
K_{\bm a}({\bf x},{\bf x}_{u}) = \text{exp}\left[-\sum_{j=1}^d  \frac{(x_{j}-x_{uj})^2}{\sigma_j^2} \right],
$$
where $\sigma_j$ is controlled by the unconstrained parameter $a_j$. In particular, for $k\neq j$, $\kappa_{u}^{jk}$ simplifies to
$$
    \kappa_{u}^{jk}
    =\kappa_{u}^{k}\left(L_{u}^{j}-\sum_{f}\kappa_{f}L_{f}^{j}\right)-\kappa_{u}\sum_{f}\kappa_{f}^{k}L_{f}^{j}.
$$
 Similarly, for $j\neq k\land j\neq l\land k\neq l$, $\kappa_{u}^{jkl}$ simplifies to
\begin{equation}\label{eq:supp_alpha_jkl}
    \kappa_{u}^{jkl} = \kappa_{u}^{kl}\left(L_{u}^{j}-\sum_{f}\kappa_{f}L_{f}^{j}\right)-\kappa_{u}^{k}\sum_{f}\kappa_{f}^{l}L_{f}^{j}-\kappa_{u}^{l}\sum_{f}\kappa_{f}^{k}L_{f}^{j}-\kappa_{u}\sum_{f}\kappa_{f}^{kl}L_{f}^{j},
\end{equation}
while, when $j=k\neq l$ or $j=l\neq k$, we need to add 
\begin{align*}
    \kappa_{u}^{l}\left(L_{u}^{jj}-\sum_{f}\kappa_{f}L_{f}^{jj}\right)-\kappa_{u}\sum_{f}\kappa_{f}^{l}L_{f}^{jj},
    && \text{or}&&
    \kappa_{u}^{k}\left(L_{u}^{jj}-\sum_{f}\kappa_{f}L_{f}^{jj}\right)-\kappa_{u}\sum_{f}\kappa_{f}^{k}L_{f}^{jj},
\end{align*}
to  \eqref{eq:supp_alpha_jkl}. If $j=k=l$, we add both of the above terms, as well as 
\begin{equation*}
\kappa_{u}\left(L_{u}^{jjj}-\sum_{f}\kappa_{f}L_{f}^{jjj}\right).
\end{equation*}

Under the unconstrained parametrisation $a_j = -\log \sigma_j$, the log-kernel is 
$$
L_{\bm a}({\bf x},{\bf x}_{u}) = -\sum_{j=1}^d (x_{j}-x_{uj})^2e^{2a_j},
$$
and its derivates w.r.t. $a_j$ are
\begin{align*}
    L^{j}=2L,
    && 
    L^{jj}=2L^{j},
    &&
    L^{jjj}=2L^{jj},
\end{align*}
and so forth, while all the mixed derivatives w.r.t. $a_{j}$ and $a_{k}$ are equal to zero.

\subsection{Derivatives of scaling penalty with respect to {\boldmath $a$}}
Section~\ref{app:scaling_penalty} introduced a penalty $q_{\tilde{s}}({\bm \zeta})$ on the observed variance of the nested effect $\tilde{s}$. This approach addresses the scaling problem described in Section~\ref{sec:scaling_issues} by ensuring that the empirical variance $\hat{\var}(\tilde{\bm s})$ of $\tilde{\bm s}$ is fixed to $c$. Below, we provide the derivatives of this penalty w.r.t. the transformations parameters, $\bm a$, for both general nested effects and a single index effects, the latter being computationally more efficient due the linearity of the transformation.

\subsubsection{General case}
Recall that the penalty defined in Section~\ref{app:scaling_penalty} is
\begin{equation*}
    q_{\tilde{s}'}= \{\hat{\var}[\tilde{s}'(\mathbf{x})] - c\}^2.
\end{equation*}
where $\hat{\var}[\tilde{s}'(\mathbf{x})]$ represents the sample variance of $\tilde{s}'(\mathbf{x})$, and $c>0$ is a constant. While $\tilde{s}'$ is a centred and scaled version of $\tilde{s}$, see \eqref{eq:scaled_nested}, the formulas provided here apply even to uncentred/unscaled transformations. Hence, below we just refer to $\tilde{s}$, but the formulas for $\tilde{s}'$ are obtained by simply plugging $\tilde{s}'$ in place of $\tilde{s}$. 

As in the previous sections, let $f^j$ represent the partial derivative of a function $f$ with respect to the $j$-th parameter. Higher-order derivatives are indicated with additional superscript letters.  Consequently, $q_{\tilde{s}}^j$ is given by
\begin{align*}
    q_{\tilde{s}}^j  = \pdif{q_{\tilde{s}}}{a_j}
    &=
    \frac{4}{n} \left[\hat{\var}(\tilde{\bm s})-c\right] \sum_{i=1}^n \left(\tilde{s}_{i}- \overline{\tilde{s}}\right)\left(\tilde{s}_i^{j}-\overline{\tilde{s}}^{j}\right)\\
    &= \frac{4}{n} \left[\hat{\var}(\tilde{\bm s})-c\right] \sum_{i=1}^n \tilde{s}_{ci}\tilde{s}_{ci}^{j}
\end{align*}
or in vector form
\begin{equation*}
    \nabla_{\bm{a}}q_{\tilde{s}}=\frac{4}{n}(\hat{\var}(\tilde{\bm{s}})-c)\bar{{\bf U}}\ts\tilde{\bm{s}}_{c}
\end{equation*}
where $\bar{{\bf U}}=\nabla_{\bm a}\ts\tilde{\bm s}-{\bf 1}{\bf 1}\ts\nabla_{\bm a}\ts\tilde{\bm s}$, with ${\bf 1}=\{1,\dots,1\}$, 
denotes the column-centred version of the Jacobian $\nabla_{\bm a}\ts\tilde{\bm s}$, and $\tilde{s}_{ci}=\tilde{s}_{i}-\overline{\tilde{s}} = \tilde{s}_i - n^{-1}{\bf 1}\ts\tilde{\bm s}$ represents the $i$-th element of centred nested effect (similar notation is adopted for higher-order derivatives). The $jk$ element of the Hessian matrix is
\begin{equation*}
    q_{\tilde{s}}^{jk}
    =\frac{8}{n^{2}}\sum_{i=1}^n \tilde{s}_{ci}\tilde{s}_{ci}^{j} \sum_{i=1}^n\tilde{s}_{ci}\tilde{s}_{ci}^{k}
    +\frac{4}{n}\left[\hat{\var}(\tilde{\bm{s}})-c\right]
    \left(\sum_{i=1}^n  \tilde{s}_{ci}^{k} \tilde{s}_{ci}^{j}
    +\sum_{i=1}^n \tilde{s}_{ci}\tilde{s}_{ci}^{jk}\right).
\end{equation*}
In matrix form
\begin{equation*}
    \nabla_{\bm{a}}\ts\nabla_{\bm{a}} q_{\tilde{s}}=\frac{4}{n}\left[\frac{2}{n}\bar{{\bf U}}\ts\tilde{\bm{s}}_c\tilde{\bm{s}}_c\ts\bar{{\bf U}}+[\hat{\var}(\tilde{\bm{s}})-c](\bar{{\bf U}}\ts\bar{{\bf U}}+{\bf P})\right],
\end{equation*}
where ${\bf P}=\sum_{i=1}^n\tilde{s}_{ci}(\nabla_{\bm{a}}\ts\nabla_{\bm{a}}\tilde{s}_{ci})$.
The third derivative array is
\begin{align*}
    q_{\tilde{s}}^{jkl}
    &=\frac{8}{n^{2}}\left[
    \sum_{i=1}^n \left(\tilde{s}_{ci}^{l}\tilde{s}_{ci}^{j}+\tilde{s}_{ci}\tilde{s}_{ci}^{jl}\right) \sum_{i=1}^n\tilde{s}_{ci}\tilde{s}_{ci}^{k} + \sum_{i=1}^n\tilde{s}_{ci}\tilde{s}_{ci}^{j} \sum_{i=1}^n\left(\tilde{s}_{ci}^{l}\tilde{s}_{ci}^{k} + \tilde{s}_{ci}\tilde{s}_{ci}^{kl}\right) 
    \right] +\\
    &\quad + \frac{4}{n} \left\{ \frac{2}{n}\sum_{i=1}^n\tilde{s}_{ci}\tilde{s}_{ci}^l \left(\sum_{i=1}^n\tilde{s}_{ci}^k\tilde{s}_{ci}^j + \sum_{i=1}^n\tilde{s}_{ci}\tilde{s}_{ci}^{jk}\right) + \right.\\
    &\qquad \qquad \left. 
    + [\hat{\var}(\tilde{\bm s})-c] \left( \sum_{i=1}^n \tilde{s}_{ci}^{kl}\tilde{s}_{ci}^j+\sum_{i=1}^n\tilde{s}_{ci}^k\tilde{s}_{ci}^{jl}+\sum_{i=1}^n\tilde{s}_{ci}^l\tilde{s}_{ci}^{jk}+\sum_{i=1}^n\tilde{s}_{ci}\tilde{s}_{ci}^{jkl}\right)\right\}.
\end{align*}
Defining ${\bf q}=\bar{{\bf U}}\ts\tilde{\bm s}_{c}$, ${\bf C}=\bar{{\bf U}}\ts\bar{{\bf U}}$
and the three-dimensional arrays ${\bf A}$ and ${\bf B}$ such that $A_{abc}=q_{a}(C_{bc}+P_{bc})$
and $B_{abc}=\sum_{i=1}^n\tilde{s}_{ci}^{a}\tilde{s}_{ci}^{bc}$, the aforementioned expression can be reformulated as
\begin{equation*}
    q_{\tilde{s}}^{jkl} 
    =\frac{8}{n^{2}} \left(A_{jkl}+A_{kjl}+A_{ljk}\right)
    +\frac{4}{n}[\hat{\var}(\tilde{\bm s})-c] \left(B_{jkl}+B_{kjl}+B_{ljk}+\sum_{i=1}^n \tilde{s}_{ci}\tilde{s}_{ci}^{jkl}\right).
\end{equation*}
Note that $A_{abc}$ and $B_{abc}$ are symmetric in $b$ and $c$ but
not in $a$ (i.e., $A_{abc}=A_{acb}$ but $A_{abc}\neq A_{cba}$).

Finally, the derivative of $\nabla_{\bm{a}}\ts\nabla_{\bm{a}} q_{\tilde{s}}$
w.r.t. a log-smoothing parameter $\rho$ is required for LAML maximisation. This is
\begin{align*}
    \dif{q_{\tilde{s}}^{jk}}{\rho} 
    &= \sum_{l=1}^d q_{\tilde{s}}^{jkl} \dif{\hat{a}_l}{\rho}\\
    &=\frac{8}{n^{2}}\left[({\bf C}+{\bf P})\hat{\bm a}_{\rho}{\bf q}\ts+{\bf q}\hat{\bm a}_{\rho}\ts({\bf C}+{\bf P})+{\bf q}\ts\hat{\bm a}_{\rho}({\bf C}+{\bf P})\right]+\\
    &\qquad +
    \frac{4}{n}[\hat{\var}(\tilde{\bm s})-c]\left[{\bf Z}\ts{\bf U}+{\bf Z}{\bf U}\ts + \sum_{i=1}^n({\bf U}\hat{\bm a}_{\rho})_{i}\tilde{s}_{ci}^{jk}+\sum_{i=1}^n\tilde{s}_{ci}\sum_{l=1}^d\tilde{s}_{ci}^{jkl}(\hat{\bm a}_{\rho})_{l}\right],
\end{align*}
where $(\hat{\bm a}_{\rho})_{l}=\dif{\hat{a}_{l}}{\rho}$ and ${\bf Z}$ is
such that $Z_{ij}=\sum_{l=1}^d\tilde{s}_{ci}^{jl}\dif{\hat{a}_{l}}{\rho}$.

\subsubsection{Single index case}
For the single index case, the linearity of the effect results in some terms being zero, allowing more efficient computation. The derivatives of $q_{\tilde{s}}$ are 
\begin{align*}
    q_{\tilde{s}}^{j}
    &=4\left[\hat{\var}(\tilde{{\bf X}}\bm{a})-c\right]\hat{\bm \sigma}_{j}\ts\bm{a},\\
    q_{\tilde{s}}^{jk}
    &=8\hat{\bm \sigma}_{j}\ts\bm{a}\bm{a}\ts\hat{\bm \sigma}_{k} + 4\left[\hat{\var}(\tilde{{\bf X}}\bm{a})-c\right]\hat{\sigma}_{jk},\\
    q_{\tilde{s}}^{jkl}
    &= 8\left(\hat{\sigma}_{jl}\bm{a}\ts\hat{\bm \sigma}_{k}+\hat{\bm \sigma}_{j}\ts\bm{a}\hat{\sigma}_{kl}\right)
    +8\hat{\sigma}_{jk}\hat{\bm \sigma}_{l}\ts\bm{a}\\
    &=8\left[\hat{\sigma}_{jl}(\hat{\bm{\Sigma}}\bm{a})_{k} 
    +\hat{\sigma}_{kl}(\hat{\bm{\Sigma}}\bm{a})_{j}
    +\hat{\sigma}_{jk}(\hat{\bm{\Sigma}}\bm{a})_{l}\right],
\end{align*}
where ${\bf X}$ is the matrix of covariates for the single-index term and $\hat{\bm \sigma}_{j}$ represents the $j$-th column or row of $\hat{\bm{\Sigma}}=\hat{\text{cov}}(\tilde{{\bf X}})$, which is the empirical maximum likelihood covariance matrix estimator (i.e., the maximum likelihood version which divides by $n$, not $n-1$). The gradient and Hessian in matrix form are
\begin{align*}
    \nabla_{\bm a} q_{\tilde{s}}
    &=2\left[\hat{\var}(\tilde{{\bf X}}\bm{a})-c\right]\nabla_{\bm a}\hat{\var}(\tilde{{\bf X}}\bm{a})\\
    &=4\left[\hat{\text{var}}(\tilde{{\bf X}}\bm{a})-c\right]\hat{\bm{\Sigma}}\bm{a},\\
    \nabla_{\bm a}\ts\nabla _{\bm a}q_{\tilde{s}}
    &=8\hat{\bm{\Sigma}}\bm{a}\bm{a}\ts\hat{\bm{\Sigma}}+4\left[\hat{\var} (\tilde{{\bf X}}\bm{a})-c\right]\hat{\bm{\Sigma}}.
\end{align*}
Additionally, we need the derivative of $\nabla_{\bm{a}}\ts\nabla_{\bm{a}}q_{\tilde{s}}$
w.r.t. $\rho$ for LAML maximisation. This is
\begin{align*}
    \dif{q_{\tilde{s}}^{jk}}{\rho} 
    &= \sum_{l=1}^dq_{\tilde{s}}^{jkl} \dif{\hat{a}_l}{\rho}
    = \sum_{l=1}^d \left[8(\hat{\sigma}_{jl} \hat{\bm{a}}\ts \hat{\bm \sigma}_{k} + \hat{\bm \sigma}_{j}\ts\hat{\bm{a}}\hat{\sigma}_{kl})
    +8\hat{\sigma}_{jk}\hat{\bm \sigma}_{l}\ts\hat{\bm{a}}\right] \dif{\hat{a}_l}{\rho},
\end{align*}
which can be computed efficiently by doing 
\begin{equation*}
    \dif{\nabla_{\bm{a}}\ts\nabla_{\bm{a}} q_{\tilde{s}}}{\rho}
    =8\left[\hat{\bm{\Sigma}}\bm{a}\otimes\left(\hat{\bm{\Sigma}}\dif{\hat{\bm{a}}}{\rho}\right)\ts
    +\hat{\bm{\Sigma}}\dif{\hat{\bm{a}}}{\rho}\otimes(\hat{\bm{\Sigma}}\bm{a})\ts
    +\hat{\bm{\Sigma}}\bm{a}\ts\hat{\bm{\Sigma}}\dif{\hat{\bm{a}}}{\rho}\right].
\end{equation*}

\section{House prices data}\label{sec:source of data}
In this section, we provide additional details on the covariate definitions and original data sources used to model house prices in London.

\begin{figure}\centering
    \includegraphics[width = 0.6\linewidth]{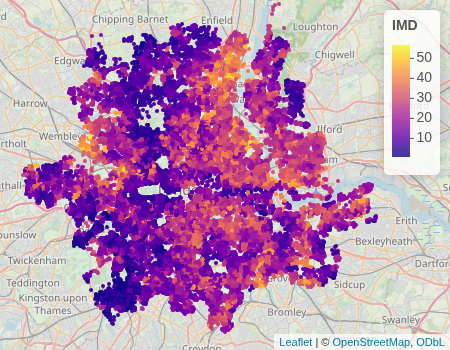}
    \caption{Map of the IMD value for each LSOA in London.} \label{fig:IMD_map}
\end{figure}

\begin{figure}\centering
    \includegraphics[width = 0.6\linewidth]{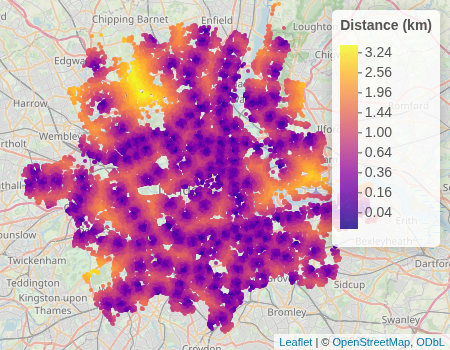}
    \caption{Distance (in km) of each sold property from the closest underground station. The colour scale is square-rooted.} \label{fig:tube_map}
\end{figure}

\begin{itemize}
    \item \textbf{Price Paid Data} is a large data set that contains details on all property transactions in England and Wales that were sold for a value and officially recorded by the HM Land Registry. It is available for download from \href{https://www.gov.uk/government/statistical-data-sets/price-paid-data-downloads}{HM Land Registry\footnote{Contains HM Land Registry data \textsuperscript{\textcopyright} Crown copyright and database right 2021. This data is licensed under the Open Government Licence v3.0.}}. The data analysed here were obtained by selecting records from the 2022 dataset where \texttt{city == "LONDON"}. Furthermore, we concentrated on observations ranging from 100,000 to 10 million pounds. The variables taken into account are:
    \begin{itemize}
    \item \textbf{Postcode}: The postcode at the time of the initial transaction;
    \item \textbf{County};
    \item \textbf{Price}: the sale price listed on the transfer deed;
    \item \textbf{Property Type}: D = Detached, S = Semi-Detached, T = Terraced, F = Flats/Maisonettes, O = Other;
    \item \textbf{Old/New}: Specifies whether the property is newly built (Y) or an older, established building (N);
    \item \textbf{Duration}: refers to the type of tenure: F = Freehold, L = Leasehold;
    \item \textbf{PPD Category Type}: Indicates the specific category of the price-paid transaction, A = standard price-paid entry, which includes single residential properties sold for value. B = Additional Price Paid entry, such as transfers under a power of sale / possessions, buy-to-let (if identified by a mortgage), transfers to non-private individuals, and sales where the property type is classified as ``Other".
\end{itemize}
\item \textbf{Data on postcode geographical locations} can be downloaded from \href{https://www.ordnancesurvey.co.uk/products/code-point-open}{Ordnance Survey National Geographic Database (OS NGD)}. 
\item \textbf{Data on best fit between postcode and Lower Layer Super Output Areas (LSOA)} can be downloaded from \href{https://geoportal.statistics.gov.uk/datasets/9c5ebee4163d435aa4defdaf348ba3c2/about}{the Office for National Statistics}. 
\item \textbf{Index of deprivation 2019} is an index that measures relative deprivation in LSOA, see Figure \ref{fig:IMD_map}. It can be downloaded from \href{https://www.gov.uk/government/statistics/english-indices-of-deprivation-2019}{the UK government data.}
\item \textbf{London Underground station data} was sourced from the GitHub repository, available at \href{https://github.com/oobrien/vis/blob/master/tubecreature/data/nr_stations.json}{https://github.com/oobrien}. The distance (in km) of each sold property from the closest underground station is shown in Figure \ref{fig:tube_map}.
\end{itemize}

